%% file: paper.tex
\documentclass[acmtog]{acmart}
\acmSubmissionID{172}

\input{header}

\graphicspath{{figures_arxiv/}}

\acmJournal{TOG}
\acmVolume{39}
\acmNumber{6}
\acmArticle{262}
\acmYear{2020}
\acmMonth{12}

\setcopyright{acmcopyright}

\acmDOI{10.1145/3414685.3417783}

\begin{document}
\title{DeformSyncNet: Deformation Transfer via Synchronized Shape Deformation Spaces}

\author{Minhyuk Sung}
\affiliation{\institution{Adobe Research}}
\authornote{This work was equally contributed by M.Sung and Z. Jiang as co-first authors.}
\author{Zhenyu Jiang}
\affiliation{\institution{The University of Texas at Austin}}
\authornote{This work was done when Z. Jiang attended Tsinghua University.}
\author{Panos Achlioptas}
\affiliation{\institution{Stanford University}}
\author{Niloy J. Mitra}
\affiliation{\institution{University College London and Adobe Research}}
\author{Leonidas J. Guibas}
\affiliation{\institution{Stanford University}}

\renewcommand\shortauthors{Sung et al.}

\input{sections/abstract.tex}

%
%


%
%


\begin{teaserfigure}
\centering
\includegraphics[width=\textwidth]{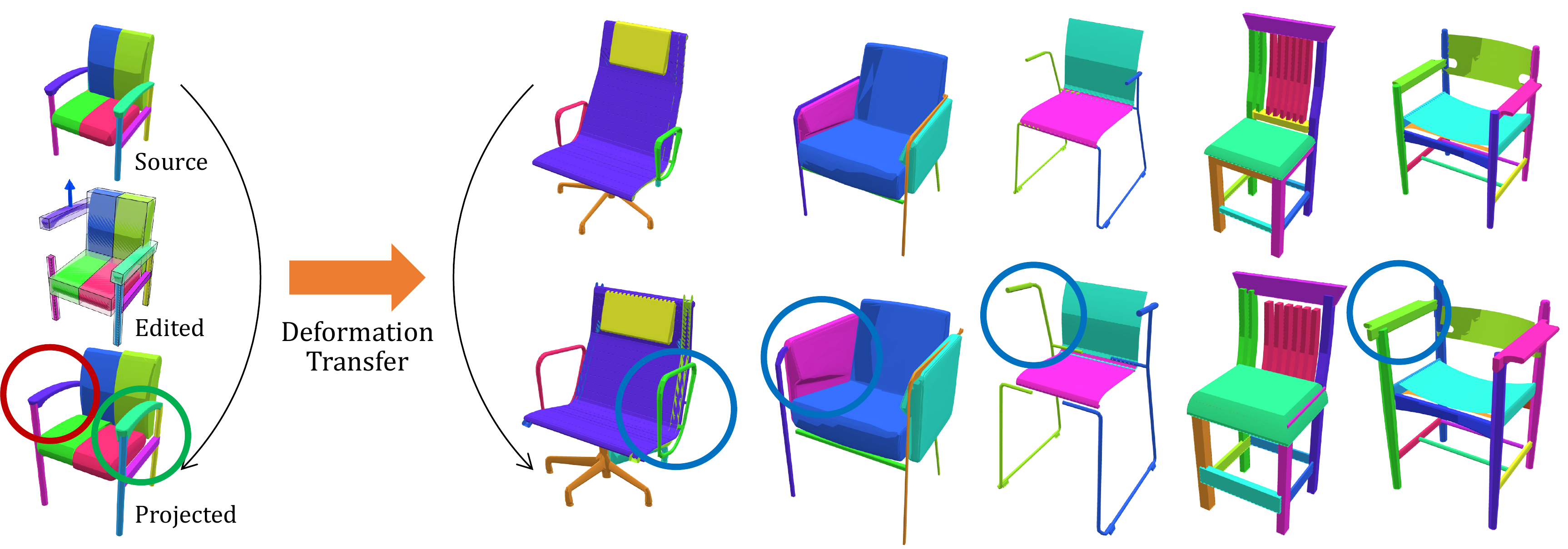}
\vspace{-\baselineskip}
\caption{\rev{
We propose \DeformSyncNet~to jointly learn an idealized canonical latent space, encoding all possible deformations from one shape to any other, as well as an individualized linear deformation space, realized in a particular way for each specific shape. These individual deformation spaces are \emph{synchronized} via the connection to the canonical space, resulting in consistent deformations across an entire shape category. Here, a user's manipulations on a shape is projected to a learned plausible shape space (left) while maintaining the shape structure such as part connectivity (red circle) and symmetry (green circle) and also instantly transferred to multiple other shapes from the same category (right), \textit{without} leveraging any correspondence information during network training.
Note how the edit only affects the other chairs with arms (blue circle). 
}
}
\label{fig:teaser}
\end{teaserfigure}

\maketitle

\input{sections/introduction}
\input{sections/related_work}
\input{sections/method}
\input{sections/results}
\input{sections/conclusion}

\revfinal{
\section*{Acknowledgments}
We thank the anonymous reviewers for their comments and suggestions.
N. J. Mitra acknowledges the support of ERC PoC Grant, Google Faculty Award, Royal Society Advanced Newton Fellowship, and gifts from Adobe. L. J. Guibas acknowledges the support of a Vannevar Bush Faculty Fellowship, a Samsung GRO grant, a Google Daydream Research Award, and gifts from the Adobe, Autodesk, and Snap corporations.
}

\bibliographystyle{ACM-Reference-Format}
\bibliography{paper}

\end{document}

%% file: header.tex
\usepackage{amsbsy,amscd,amsfonts,amsgen,amsmath,amstext,amsopn,amssymb}
\usepackage{array}
\usepackage{booktabs}
\usepackage{color}
\usepackage{graphicx}
\usepackage{float}
\usepackage{mathtools}
\usepackage{multirow}
\usepackage{enumerate}
\usepackage{tabularx}
\usepackage{tikz-cd}
\usepackage{wrapfig}
\usepackage{xcolor}
\usepackage{xspace}

\newcolumntype{L}{>{\raggedright\arraybackslash}X}
\newcolumntype{C}{>{\centering\arraybackslash}X}

\DeclareMathOperator*{\argmin}{argmin}

\makeatletter
\DeclareRobustCommand\onedot{\futurelet\@let@token\@onedot}
\def\@onedot{\ifx\@let@token.\else.\null\fi\xspace}
 
\def\ie{\emph{i.e}\onedot}

\def\etal{\emph{et al}\onedot}

\makeatother

\newcommand{\DeformSyncNet}[0]{\textsc{DeformSyncNet}}
\newcommand{\DSN}[0]{\textsc{DSN}}

\newcommand{\rev}[1]{{#1}}
\newcommand{\revfinal}[1]{#1}

%% file: sections/abstract.tex
\begin{abstract}
    
Shape deformation is an important component in any geometry processing toolbox. The goal is to enable intuitive deformations of single or multiple shapes or to transfer example deformations to new shapes while preserving the plausibility of the deformed shape(s). Existing approaches assume access to point-level or part-level correspondence or establish them in a preprocessing phase, thus limiting the scope and generality of such approaches. 
We propose \DeformSyncNet, a new approach that allows consistent and synchronized shape deformations without requiring explicit correspondence information.
Technically, we achieve this by encoding deformations into a class-specific idealized latent space while decoding them into an individual, model-specific linear deformation action space, operating \textit{directly} in 3D. The underlying encoding and decoding are performed by specialized (jointly trained) neural networks. By design, the inductive bias of our networks results in a deformation space with several desirable properties, such as path invariance across different deformation pathways, which are then also approximately preserved in real space.
%
We qualitatively and quantitatively evaluate our framework against multiple alternative approaches and demonstrate improved performance. 
\end{abstract}

%% file: sections/introduction.tex

\section{Introduction}
\label{sec:introduction}

Shape deformation is an essential task for both computer graphics and computer vision. For example, in shape retrieval applications, one can deform a retrieved shape to better meet the user's desiderata. No matter how large the available 3D shape repositories have become~[\citeauthor{Warehouse,TurboSquid,GrabCAD}], shape deformation still plays a critical role in shape modeling and shape retrieval because there are infinite imaginable shape variations, and creating novel shapes from scratch is still a taxing task.
Due to its importance and broad impact, shape deformation has been extensively studied by the geometry processing community in the last decades. Such work can be broadly grouped into two categories. In the first category, one wants to plausibly deform \emph{a single shape based on user input/interactions}~\cite{Sorkine:2004,Igarashi:2005,Lipman:2005,Yumer:2014}. Specifically, one can allow users to edit shapes using a set of \emph{deformation handles}, which are intuitive and procedural parameters either directly coming from the 3D CAD models or generated from parametrizing or approximating each part of a 3D model~\cite{Xu:2009,Zheng:2011,Mo:2019:StructureNet}. Such deformation handles, however, mostly overparametrize the deformation space making it hard to maintain the plausibility or semantic constraints of a shape under arbitrary parameter changes. Consequently, user edits have to be constantly \emph{projected} to the plausible deformation space \cite{snapIt,Gal:2009}, which is an error-prone process without strict quality guarantees. In the second category, one may want to apply \textit{deformation transfer}, where algorithmic techniques propagate the deformation prescribed on one shape to another shape, or even to a collection of shapes~\cite{Sumner:2004,Zhou:2010,Ben-Chen:2009,Chen:2010,Baran:2009,Ovsjanikov:2011,Fish:2014}. While deformation transfer alleviates the burden of individually deforming each shape, such algorithms typically expect explicit correspondences between shapes or deformation handles, which is an unrealistic demand in practice.

Nevertheless and leaving for a moment aside the practical considerations, if one assumes the existence of correspondences between deformation handles, some natural ideas emerge. In the projection's case, the correspondences can help us discover and exploit statistical \emph{correlations} among the deformation parameters, which can be further incorporated in, and improve the quality of the projection~\cite{Fish:2014}. In the case of deformation-transfer, one can simply transfer the desired changes at the granularity of handles (i.e., from the source's handle(s) to the target's ones) tapping on the correspondence-induced regularization. These are some of the important reasons of why many previous works assume that correspondences are available as groundtruth supervision~\cite{Sumner:2004,Zhou:2010,Yang:2018,Ben-Chen:2009,Chen:2010,Baran:2009}, or estimate correspondences in a preprocessing phase~\cite{Yumer:2014}. 

In practice, however, computing dense correspondences is expensive or even ill-defined for heterogeneous collection of shapes, e.g., a collection of chairs that includes swivel and four-legged models. Some 3D datasets provide part-level semantic annotations~\cite{Yi:2016,Mo:2019:PartNet}, but these are insufficient to indicate correspondences for \emph{fine-grained} deformations. For all these reasons, in this work, we propose novel neural networks that allow us to do deformation projection and transfer without relying on \emph{any} precomputed notion of shape correspondences (i.e., not even between handles).

Concretely, we introduce \DeformSyncNet, a neural architecture that jointly learns (i)~an \emph{idealized canonical latent space} of shape encodings, where all possible deformations from any shape to any other are possible, as well as (ii)~\emph{individual} linear deformation spaces for each real shape that reflect the particular ways that the shape may (\textit{or may not} be able to) deform in 3D. These shape-specific deformation spaces are synchronized by being connected through the canonical space. We design the canonical latent space as an \emph{affine} space where shape deformations, arising as vectors connecting latent shape encodings, are the main objects of interest --- and have a meaning irrespective of the source and target shapes involved. A deformation is transferred by decoding that latent deformation vector in the context of the new source shape and applying the resulting 3D deformation to that shape in real space.
The individual deformation spaces, each of which is represented as a \emph{dictionary} of linear deformations, are linked together so as to encourage the network to share the same parameters for the individual linear functions. Additionally, the properties of the affine space naturally structure the latent space to impose cycle consistency between various deformations, without additional loss functions or regularization during the neural network training. 

Our approach can benefit from (but does not require) deformation handles. If we are given deformation handles for a particular shape, these can be easily connected to the individual deformation space we learn. Since, both spaces are linear in our setting, the projection from editing via deformation handles to a plausible shape can be simply computed as an orthogonal projection to a linear learned space. During network training, we also enforce that the learned deformation space for each shape becomes a \emph{subspace} of the given deformation handle space so that it follows the given deformation operations, while capturing common constraints across the shapes. In this way, \emph{correlations} among the deformation handles can emerge in the learned space.

In the experiments, we successfully apply our framework for the purposes of \emph{co-editing} of human-made shapes and show qualitatively intuitive results. Quantitatively, we show that our trained networks outperform other modern (neural-net-based) and classical (ICP-based) shape deformation approaches, both in terms of fitting accuracy and quality of deformation transfer.

In summary, our contributions are:
\begin{enumerate}[(i)]
    \item a novel approach for learning synchronized linear deformation spaces for each shape in a category, without explicit correspondences; 
    \item a unified framework enabling both projection of user-edited shapes to plausible counterparts in the shape space as well as transfer of the deformation to other shapes in the same category; and
    \item a neural network design that is simple yet outperforms existing deep deformation methods in terms of both quality of fitting and deformation transfer.
\end{enumerate}

\clearpage


%% file: sections/related_work.tex

\section{Related Work}
\label{sec:related_work}

\subsection{3D Shape Deformation}
3D shape deformation has been a long-standing problem in computer graphics. Earlier work introduced methods enabling interactive free-form deformations, while preserving local characteristics of the shape. Some well-known examples are Laplacian editing~\cite{Sorkine:2004} and as-rigid-as-possible manipulation~\cite{Igarashi:2005} that regularize deformations to maintain local curvature based on mesh Laplacians and local rigidity, respectively. Recent work focuses more on target-driven deformation, deforming and \emph{fitting} a source shape to a target. For the fitting stage, researchers have used iterated closest point~(ICP) iterations~\cite{Li:2008,Huang:2017} to establish point-wise correspondences and subsequently minimized an energy function based on them. However, ICP is prone to fail when the source shape is significantly different from the target, violating the ICP locality assumption. Recent neural-net-based learning approaches directly predict the deformation offset either in voxel grid~\cite{Yumer:2016,Jack:2018,Kurenkov:2018,Hanocka:2018} or for point samples on the source shape~\cite{3DN,CycleConsistency,Mehr:2019} resulting in better fitting accuracy.

Both traditional and recent learning-based approaches do not, however, learn the shape variability from the given dataset, and thus cannot verify the semantic \emph{plausibility} of the deformed shapes. Specifically, they cannot examine whether the deformed shape looks like one of the exemplars in the database. Instead, we train a network by performing target-driven deformation, while simultaneously requiring it to learn the shape variation space. Hence, in interactive editing, we can project the user's input through the deformation handles to the learned shape space and preserve semantic plausibility. We also demonstrate in our experiments that our method performs the target-driven deformation comparably or even better than previous methods.

Note that our method also differs from other deep 3D generative models that learn a shape variation space~\cite{Wu:2016,Achlioptas:2018} in that it does not generate shapes from a latent representation. Instead, when decoding, it takes an existing shape as input and generates a variant by deforming it directly in 3D. Hence, our method enables \emph{reusing} existing 3D shape data with their associated meta-information, e.g., mesh structure, color, texture, and part hierarchy --- which can be carried along in the deformation. 

\subsection{Deformation Transfer}

Deformation transfer has remained an important problem in 3D shape editing after Sumner and Popovic~\shortcite{Sumner:2004} introduced the concept in their pioneering work. The goal is to automatically propagate the result of shape deformation performed on one shape to others, thus saving users' time and effort. This problem has been comprehensively investigated in many previous works. While Sumner and Popovic~\shortcite{Sumner:2004} require dense shape correspondences, Zhou~\etal~\shortcite{Zhou:2010} and Yang~\etal~\shortcite{Yang:2018} extended their work to only use keypoint correspondences. Ben-Chen~\etal~\shortcite{Ben-Chen:2009} and Chen~\etal~\shortcite{Chen:2010} overcame the limitation of requiring single-component manifold meshes, and proposed cage-based methods to deal with any representations of shapes. Baran~\etal~\shortcite{Baran:2009} improved the method to cope with very different shapes for the transfer (e.g., humans to animals) and require only region-level semantic correspondences.

The main remaining limitation is the necessity of shape correspondences as input (in the level of either points, parts, or cages). Recently, Gao~\etal~\shortcite{Gao:2018} proposed a neural-net-based method that enables inter-class deformation transfer without correspondences across the classes. Their network, however, still needs to have \emph{intra-class} correspondences of shapes during the training. In our work, we aim to deal with a diverse collection of man-made 3D shapes, such as ShapeNet models~\cite{ShapeNet}, where the correspondences between shapes are not identified or even clearly defined.
Also, their work, and a recent work of Yin~\etal~\shortcite{Yin:2019}, aim to learn transform of shapes across \emph{different} domains, inspired by analogous work in the image space~\cite{Isola:2017,Zhu:2017}, whereas our goal is to learn  deformation transfer for shapes in the \emph{same} domain.

Another notable exception is the StructEdit work~\cite{Mo:2019:StructEdit} that learns transfer of shape differences without any correspondence supervision. Compared with our method, StructEdit transfers \emph{structural or topological} differences of shapes, which cannot be immediately applied to modify a mesh or a point cloud. It also strongly depends on training with shape data annotated with consistent hierarchies~\cite{Mo:2019:PartNet}. In contrast, our method focuses on \emph{continuous} deformation and direct editing of existing 3D models.

Deformation transfer is closely related to \emph{shape analogies} (originated from image analogies~\cite{Hertzmann:2001}), finding a shape x such that $a : b = c : x$ given shapes $a$, $b$, and $c$. Rustamov~\etal~\shortcite{Rustamov:2013} first introduced a method performing shape analogies based on functional correspondences. More sophisticated recent variations on shape analogies and shape generation based on functional correspondences include~\cite{huang2019limit}, \cite{Huang_2019_ICCV}. Wu~\etal~\shortcite{Wu:2016} showed how the latent space of 3D GAN can be used for that (similar to word2vec~\cite{Mikolov:2013} in machine learning). Compared with these, we neither require correspondences nor decode directly from a latent space without deforming an input shape.

Lastly, we note that similar ideas to deformation transfer have been also studied for other research topics in graphics such as style transfer~\cite{Xu:2010,Ma:2014} and motion retargetting~\cite{Xia:2015,Villegas:2018}, but under the same assumption of that shape correspondences are available.

\subsection{Shape Constraint Analysis}

While many 3D human-made objects are highly structured, deformation handles accompanied by the 3D models often do not fully acknowledge the underlying structure and thus allow breaking it with arbitrary modifications. As alternatives, researchers have introduced 3D shape analyses extracting the global structure of 3D shapes for editing, such as orthogonality and parallelism of wireframes~\cite{Gal:2009} and bounding boxes~\cite{Zheng:2011}, symmetry~\cite{Wang:2011}, and articulation of parts~\cite{Xu:2009}. Such analyses for \emph{individual} shapes are, however, not able to capture \emph{semantic} constraints or correlations of the handles that can only be observed from families of shapes. Hence, the other line of work extended these ideas to shape \emph{co-analysis} and aimed to discover the semantic relationships among the handles from a collection of shapes. For example, Kim~\etal~\shortcite{Kim:2013} fit a template part bounding box structure to shapes in order to be able to cluster shapes and find part-level correspondences. Based on such fitted template structures, Ovsjanikov~\etal~\shortcite{Ovsjanikov:2011} demonstrates ways of exploring the shape space, 
Zheng~\etal~\shortcite{Zheng:2014} investigates co-occurrence of parts, 
and Fish~\etal~\shortcite{Fish:2014} calculate joint distributions of part box bounding parameters. Yumer~\etal~\shortcite{Yumer:2014} also discovers co-constrained abstraction of shapes in the same family. These works, however, analyze shapes based on a \emph{template} structure and thus cannot be scaled to handle 3D models in online repositories that have immense diversity. 

%% file: sections/method.tex
\section{Method}

As mentioned in the introduction, key to our approach is an idealized latent space where points represent shapes, and vectors connecting points represent shape differences or deformations. A traditional latent space approach would implement deformation transfer as simply adding the deformation vector to the point representing a new latent source shape (in order to obtain its deformed target version), followed by a decoding of the latter from latent to 3D space (Figure~\ref{fig:ae}). Instead, we propose to decode the deformation vector itself into a deformation \emph{action} which can be applied \emph{directly} in real 3D space to the new shape to be deformed.

In Section~\ref{sec:deformation_space}, before we discuss the realization of these operations on real shapes, we introduce an abstract mathematical framework of how things might work
when point-wise correspondences across the shapes are known.
We then relax these idealized assumptions to generate
a specialized set of deformations a real shape can undergo,
by developing \emph{shape-specific} deformation action dictionaries.
In Section~\ref{sec:projection}, we introduce how the learned shape deformation space can be leveraged in a practical shape editing scenario to produce a plausible output shape by \emph{projecting} the user's input to the deformation space.
In Section~\ref{sec:network_design}, we describe how we implement the shared functional form of deformations via neural networks.
Finally, in section~\ref{sec:network_comparisons}, we compare our neural networks with the other recent neural networks learning deformations and discuss the advantages of our approach.

\subsection{Abstract Deformation Space}
\label{sec:deformation_space}
We consider a collection of shapes
\revfinal{$\overline{X}$, where each shape $x \in \overline{X}$ is represented as a point cloud with $n$ points (i.e., $\forall x \in \overline{X}$, $x \in \mathbb{R}^{3n}$)}.
We aim to encode deformations between all ordered pairs in \revfinal{$\overline{X}$} within a \revfinal{low-dimensional continuous} canonical latent space. In particular, we create an \emph{affine action space} $\langle X, V, \oplus  \rangle$, where \revfinal{$\overline{X} \subseteq X \subseteq \mathbf{R}^{3n}$,} $V$ is a $k$-dimensional vector space, and $\oplus$ is an action of the additive group of $V$ on the set $X$: $X \times V \rightarrow X$.
An affine action space is defined with the following three properties~\cite{Gallier:2011,Tarrida:2011}:

\begin{enumerate}[(i)]
    \item (Free action) $x \oplus \vec{0} = x$, for every $x \in X$. 
    \item (Additive action) $(x \oplus u) \oplus v = x \oplus (u + v)$, for every $x \in X$, and every $u, v \in V$.
    \item (Transitive action) For every $x, y \in X$, there exists a unique $v \in V$ such that $x \oplus v = y$.
\end{enumerate}

Note that the vector space $V$ parametrizes not the shape (point) space but the deformation (vector) space of shape differences. Our goal is to make a vector $v \in V$ be a \emph{anchor-free} representation of deformation, \emph{forgetting} the original shape where the deformation is applied. Thus, the same type of deformation action can be applied to any arbitrary shape $x \in X$ by applying the same deformation parameter $v \in V$ --- i.e., the deformation $\overrightarrow{x y} \in V$ such that $x \oplus \overrightarrow{x y} = y$ can be transferred to the other shape $z \in X$ by computing $z \oplus \overrightarrow{x y}$.

\begin{figure}[t!]
    \includegraphics[width=\columnwidth]{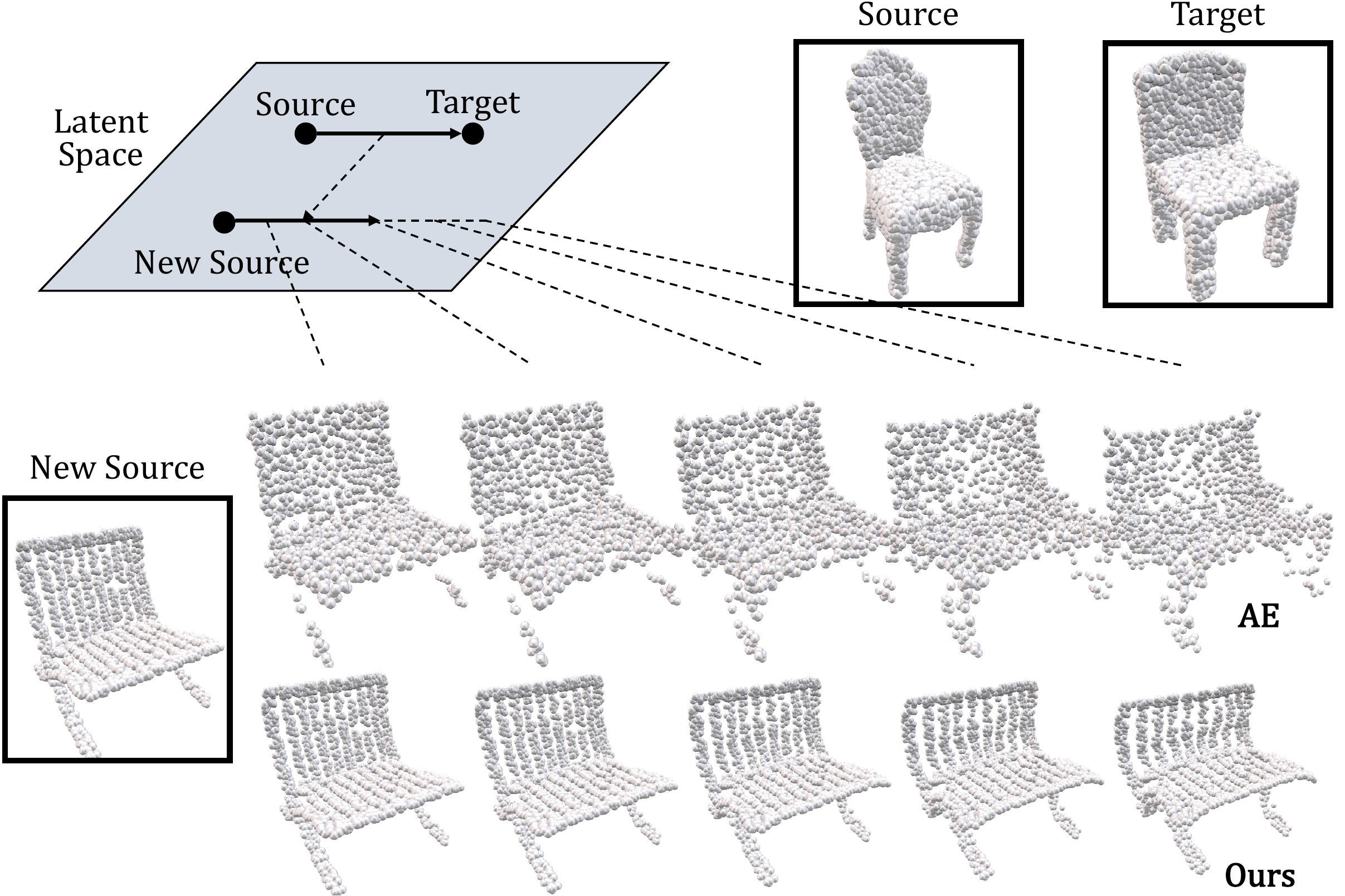}
    \vspace{-\baselineskip}
    \caption{Comparison of latent vector transfer between an autoencoder (AE) latent space~\cite{Achlioptas:2018} and our latent space. The vector from source shape to target is transferred to another source shape, and several points are sampled along the direction of the deformation vector, at scales $0.25$, $0.5$, $1.0$, $1.5$, and $2.0$. In the autoencoder latent space, the vector does not convey the \emph{difference} between two shapes and, sometimes, even pushes the new shape outside the valid region, making it less plausible. In our latent space, the difference between the source and target shapes is properly transferred, regardless of the location of the new source shape.}
    \label{fig:ae}
    \vspace{-0.5\baselineskip}
\end{figure}

We choose an affine space as our latent space of deformations because of its desirable properties, facilitating exploration of the space in downstream applications. From the basic properties above, the following properties can also be derived (Refer to Gallier~\etal~\shortcite{Gallier:2011} and Tarrida~\etal~\shortcite{Tarrida:2011} for proofs. $\overrightarrow{x y} \in V$ denotes a deformation vector from shape $x$ to $y$, i.e., $x \oplus \overrightarrow{x y} = y$.):

\begin{enumerate}[(i)]
    \item (Identity) $\,$ $\overrightarrow{x x} = \vec{0}$, for every $x \in X$. 
    \item (Anticommutativity) $\,$ $\overrightarrow{x y} = -\overrightarrow{y x}$, for every $x, y \in X$. 
    \item (Transitivity) $\,$ $\overrightarrow{x y} + \overrightarrow{y z} + \overrightarrow{z x} = \vec{0}$, for every $x, y, z \in X$. 
    \item (Parallelogram law) $\,$ $\overrightarrow{x y} = \overrightarrow{z w} \Leftrightarrow \overrightarrow{x z} = \overrightarrow{y w}$, $\forall \; x, y, z, w \in X$. 
\end{enumerate}


The challenge here is how to make the vectors in $V$ encode the \emph{same} type of deformation(s) across all the shapes in $X$.
Consider a shape autoencoder, where \revfinal{$\mathcal{E}: \mathbb{R}^{3n} \rightarrow \mathbb{R}^k$ and $\mathcal{D}: \mathbb{R}^k \rightarrow \mathbb{R}^{3n}$} are encoding and decoding neural networks, respectively.
Given these mappings, the simplest way to create an affine space for the shape differences is to take the Euclidean embedding space \revfinal{$\mathbb{R}^k$} as the vector space $V$ of the affine space: $x \oplus v \coloneqq \mathcal{D} \left( \mathcal{E}(x) + v \right)$.

A \emph{vanilla} autoencoder, however, fails to  structure the embedding space in a way that a vector indicates the same type of deformation \emph{everywhere} over the space. Figure~\ref{fig:ae} shows an example of taking a vector from the source to target shape in the embedding space and adding this to a new source shape, with different scales along the vector. This fails to make the new shape adapt the deformation from the source to target, transforming it into implausible shapes. We aim to design the latent space so that such vector addition can properly transfer the shape difference. 

\paragraph{Linear Deformation Representations}
Let us first assume that point-wise correspondences are known for all pairs of shapes --- to understand a simpler setting.
In this case, we can specify a deformation as an \emph{offset vector} for corresponding points.
When the points are ordered in a consistent way so that corresponding points across the shapes have the same index in the order, we can consider a \emph{linear} deformation function as an action $\oplus$ of the affine space:
\begin{align}
    x \oplus v \coloneqq \mathbf{A} v + x,
\end{align}
where $\mathbf{A} \in \mathbb{R}^{3n \times k}$, with $k$ being the dimension of $V$. The deformation for a $v \in V$
is now explicitly defined as adding per-point offsets $\left( \mathbf{A}v \right)$ to the source point cloud, and the same $v$ produces the same offsets for the corresponding points across all the shapes.
This action on shapes is
free ($x \oplus \vec{0} = x$), transitive
\revfinal{if $k$ is smaller or equal to $3n$, yet large enough to capture all possible differences of shapes in $X$}, and also is in the additive group of $V$ over $X$. In the context of an autoencoder, this can be understood as constructing a latent shape space to be decoded to a \emph{linear subspace} $\mathbf{A}$ of the shape point clouds, so that a free vector $v$ over that latent space describes the same point displacements in $3D$ space, regardless of the original shape.
\revfinal{Alternatively, one can simply interpret $\mathbf{A}$ as a set of principal axes of the point clouds $x \in \overline{X}$, in the context of PCA.}

Now we consider the case when the point-wise correspondences are \emph{unknown}, and possibly even \emph{ill-defined}.
\footnote{For a heterogeneous collection of shapes, particularly for human-made objects, the correspondences may not be clearly defined for some pairs, either geometrically or semantically.}
Hence, in this case, it is not possible to define a canonical linear deformation function for all shapes with a single matrix $\mathbf{A}$. Instead, we propose to predict a matrix $\mathbf{A}$ for each individual shape $x \in X$ using a neural network $\mathcal{F}: X \rightarrow \mathbb{R}^{3n \times k}$. By denoting the output for the input shape $x$ as $\mathbf{A}_x$, the action $\oplus$ is now written as follows:
\begin{align}
    x \oplus v \coloneqq \mathbf{A}_x v + x.
    \label{eq:source_dependent}
\end{align}

The action is still free and transitive \revfinal{(with a large enough $k \leq 3n$)} but is \emph{not} in the additive group anymore, since now $\mathbf{A}_x$ is source-dependent; $\mathbf{A}_x$ and $\mathbf{A}_y$ for different shapes $x$ and $y$ may be inconsistent. To impose the property the additive action above, we jointly train a shape encoder $\mathcal{E}: X \rightarrow \mathbb{R}^k$ along with the above network $\mathcal{F}$ so that the deformation vector $\overrightarrow{x y}$ from shape $x$ to $y$ is given by $\mathcal{E}(y) - \mathcal{E}(x)$, while the matrix $\mathbf{A}_x$ is predicted as $\mathcal{F}(x)$. The deformation from the point cloud $x$ to $y$ is now computed as follows:
\begin{align}
    d(x \rightarrow y) \coloneqq  \mathcal{F}(x) \left( \mathcal{E}(y) - \mathcal{E}(x) \right) + x\,,
    \label{eq:deformation}
\end{align}
where the deformation vector is computed in latent space through $\mathcal{E}$, decoded into real space by $\mathbf{A}_x$ as a set of point offsets in an $x$-specific way, and added to the original point cloud for $x$.

This utilization of the two networks $\mathcal{E}$ and $\mathcal{F}$ realizes our overall plan: learning (i)~a canonical affine deformation space from $\mathcal{E}$; and (ii)~an individual linear (affine) deformation space for each shape from $\mathcal{F}$, and connecting these individual shape action spaces by sharing the same source latent space. The individual deformation spaces, defined by the matrices $\mathbf{A}_x$, are thus \emph{synchronized} across the shapes via the connection through the canonical latent deformation space.

Empirically, even without the joint network training, the matrix $\mathbf{A}_x$ can be predicted consistently for \emph{similar} input shapes due to the characteristics of neural networks when learning a \emph{smooth} function --- several recent works have proposed unsupervised methods finding shape correspondences based on this network property~\cite{Tulsiani:2017,groueix2018papier,Sung:2018,Zhao:2019,Li:2019,Genova:2019}. However, we found that enforcing the property of the additive action regularizes the $\mathbf{A}_x$ matrices to become consistent even when the shapes are dissimilar (see Section~\ref{sec:qualitative_evaluations} and Figure~\ref{fig:dictionary}).

\paragraph{Relation to Deep Functional Dictionaries}
The matrix $\mathbf{A}_x$ can be interpreted as a \emph{dictionary} of deformations (point-wise displacements), and in this context, we call the network $\mathcal{F}$ a \emph{deformation dictionary predictor} in the rest of the paper. The idea of learning dictionaries of functions on shapes is analogous to the Deep Functional Dictionaries work by Sung~\etal~\shortcite{Sung:2018}, but there are two main differences. First, our dictionaries are learned from pairwise deformations, not from the input functions over the shapes (thus, out work is purely self-supervised).
Second, we synchronize our dictionaries by explicitly learning a mapping of the dictionaries to a canonical space, instead of solely relying on the smoothness of the learned functions.

\subsection{Deformation Projection}
\label{sec:projection}

Many 3D models in online repositories are either described with geometric primitives or spline surfaces or are decomposed into smaller parts that can be easily approximated with bounding primitives~\cite{Yi:2017,Sung:2017,Mo:2019:PartNet}. This information provides the user with intuitive deformation handles, although mostly they \emph{overparametrize} the deformation space, meaning that arbitrary changes of the parameters do not always give a valid shape. When assuming that the deformation function of the given handles is \emph{linear} --- we observed that most deformation parameters coming from the human-made 3D models are translation and scaling of parts, which are linear deformations --- in our framework, we can easily project the user input on the given handles to the learned deformation space using simple computation. Additionally, we can guide our networks during training to learn the deformations that are present in the given parameter space.

\paragraph{Projecting Shape Editing to Deformation Space}
Let the input linear deformation function (defined by the deformation handles) denoted as $\mathbf{B}_x \left( z +z_0 \right)$, where $\mathbf{B}_x \in \mathbb{R}^{3n \times m_x}$, $m_x$ is the number of deformation handles, and $z_0 \in \mathbb{R}^{m_x}$ is the default parameters. We also have the learned linear deformation function for the shape: $\mathbf{A}_x v + x$ (Equation~\ref{eq:source_dependent}), where $x=\mathbf{B}_x z_0 \in \mathbb{R}^{3n}$ is the initial shape and $\mathbf{A}_x \in \mathbb{R}^{3n \times k}$ gives a \emph{valid} variation space of $x$. The goal of the projection is this: given the user input via the deformation handles $z \in \mathbb{R}^{m_x}$, we find $v \in \mathbb{R}^k$ (a valid variation) that minimizes the difference of point-wise offsets from the edited shape to the valid shape: $\left\Vert \mathbf{B}_x z - \mathbf{A}_x v \right\Vert_2^2$.
In particular, we consider a shape editing scenario where the user edits the shape through \emph{some} of the deformation handles and prescribes values for them at each time, and then the system automatically projects the edit to the valid space. Thus, given the \emph{equality} constraints to the deformation handles $\mathbf{C} z  = z_{\text{in}}$, where $\mathbf{C} \in \mathbb{R}^{m_x \times m_x}$ is a matrix indicating the selected handles and $z_{\text{in}} \in \mathbb{R}^{m_x}$ is the sparse user inputs, we find $\widehat{z} \in \mathbb{R}^{m_x}$ defined as follows:
\begin{align}
\begin{split}
    \widehat{z} \coloneqq &\argmin_{z} \min_v \left\Vert \mathbf{B}_x z - \mathbf{A}_x v \right\Vert_2^2 \\
    & \text{s.t.} \quad \mathbf{C} z  = z_{\text{in}}\,.
\end{split}
\label{eq:projection}
\end{align}

When $\mathbf{B}_x^\prime$ denotes the matrix $\mathbf{B}_x$ except for the \emph{columns} of the selected handles, Equation~\ref{eq:projection} can be written in a following form:
\begin{align}
    \widehat{z^\prime} \coloneqq &\argmin_{z^\prime} \min_v \left\Vert \left( \mathbf{B}_x^\prime z^\prime + c \right) - \mathbf{A}_x v \right\Vert_2^2,
\label{eq:projection_rewritten_1}
\end{align}
where $\widehat{z^\prime}$ is the best values for the rest of the parameters, and $c = \mathbf{B}\mathbf{C}z_{\text{in}}$. For any point cloud $y \in \mathbb{R}^{3n}$, the projection distance to the linear space of $\mathbf{A}_x$ is computed as:
\begin{align*}
\min_v \left\Vert \mathbf{A}_x v - y \right\Vert_2^2 = \left\Vert \left( I - \mathbf{A}_x\mathbf{A}_x^\dagger \right) y \right\Vert_2^2\,,
\end{align*}
where $\mathbf{A}_x^\dagger$ is a pseudoinverse of $\mathbf{A}_x$ (note that $\mathbf{A}_x\mathbf{A}_x^\dagger \neq I$ since $\mathbf{A}_x$ is a thin matrix, i.e., $3n \gg m_x$). Hence, Equation~\ref{eq:projection_rewritten_1}, again, can be rewritten as follows:
\begin{align}
    \widehat{z^\prime} \coloneqq &\argmin_{z^\prime} \left\Vert  \left( I - \mathbf{A}_x\mathbf{A}_x^\dagger \right)  \left( \mathbf{B}_x^\prime z^\prime + c \right) \right\Vert_2^2,
\label{eq:projection_lsq}
\end{align}

The solution of this linear regression $\widehat{z^\prime}$ is defined as:
\begin{align}
    \widehat{z^\prime} &= \mathbf{P}_x^\dagger \mathbf{Q}_x c,
\end{align}
where $\mathbf{P}_x = \left( I - \mathbf{A}_x\mathbf{A}_x^\dagger \right) \mathbf{B}_x^\prime$, $\mathbf{Q}_x = -\left( I - \mathbf{A}_x\mathbf{A}_x^\dagger \right)$, and $\mathbf{P}_x^\dagger$ is pseudoinverse of $\mathbf{P}_x$. Note that, when the user edits through one deformation at each time, $\mathbf{P}_x^\dagger$ and $\mathbf{Q}_x$ can be precomputed for each handle.

Some deformation handles may have inequality constraints; e.g., scale parameters must be positive numbers. The least squares problem in Equation~\ref{eq:projection_lsq} can also be quickly solved with inequality constraints using standard techniques such as interior-point methods or active-set methods.

\paragraph{Projecting Network Output to Deformation Handle Space}

At training time, we can also project the output deformed shape of the networks to the given deformation handle space to make the deformation \emph{describable} with the given handles (the effect of this option is investigated in Section~\ref{sec:effect_of_projection}):
\begin{align}
    d_{\text{proj}}(x \rightarrow y) \coloneqq \mathbf{B}_x \mathbf{B}_x^\dagger \left( \mathcal{F}(x) \left( \mathcal{E}(y) - \mathcal{E}(x) \right) + x \right).
    \label{eq:deformation_proj}
\end{align}


We remark that our framework can easily exploit any type of linear deformation handles, even when they are inconsistent and their numbers are different across the shapes.

\subsection{Neural Network Design and Losses}
\label{sec:network_design}

\begin{figure}[!t]
    \includegraphics[width=\columnwidth]{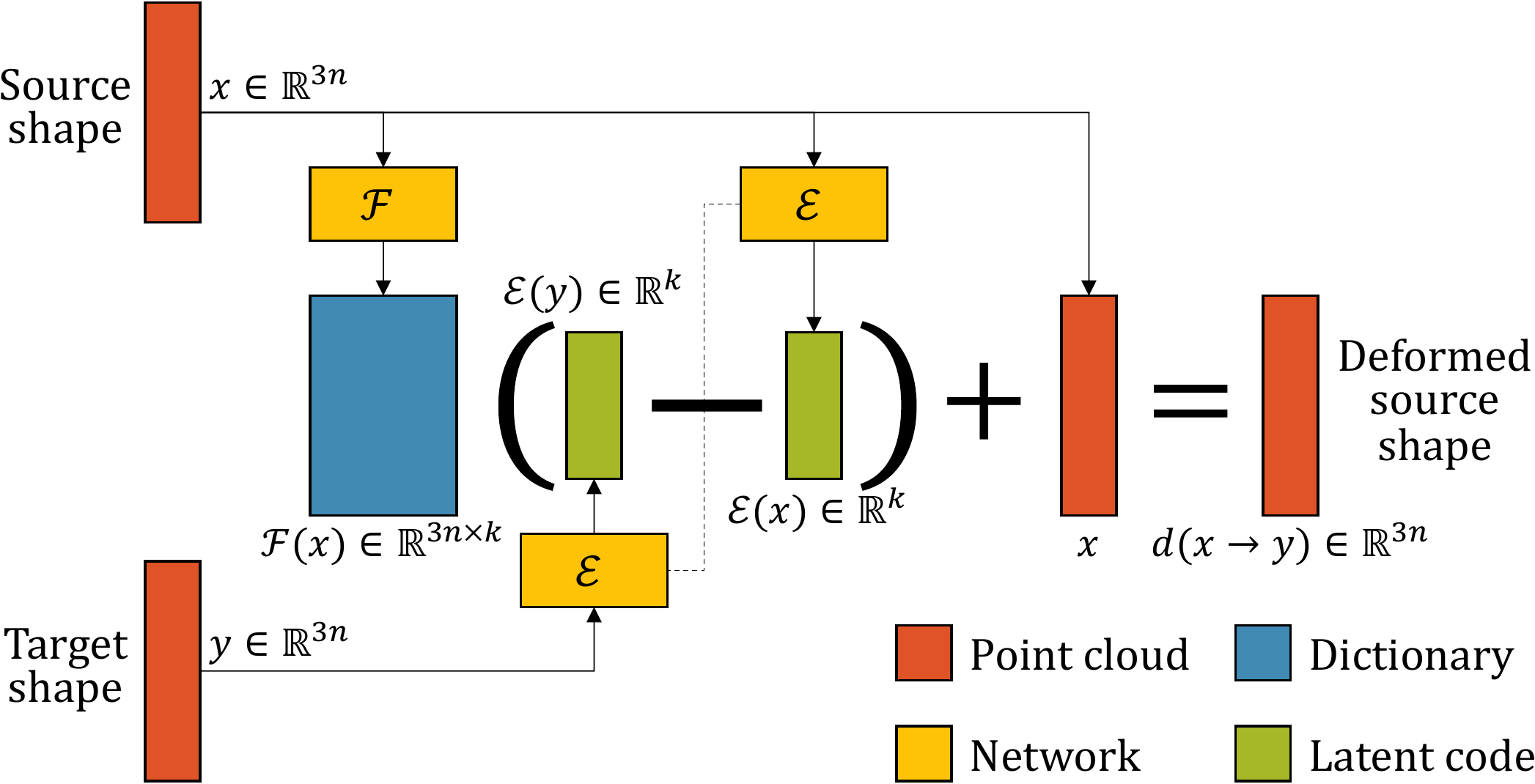}
    \vspace{-\baselineskip}
    \caption{Neural network pipeline. The encoder $\mathcal{E}$ takes both the source and target point clouds, and computes the latent deformation vector as $\left( \mathcal{E}(y) - \mathcal{E}(x) \right)$. The deformation dictionary predictor $\mathcal{F}$ takes only the source shape and predicts the dictionary $\mathcal{F}(x)$. The outputs of these two modules are multiplied and added to the source point cloud to produce the deformed shape as $ x \rightarrow x + \mathcal{F}(x)(\mathcal{E}(y)-\mathcal{E}(x))$.}
    \label{fig:pipeline}
    \vspace{-0.5\baselineskip}
\end{figure}

For the networks $\mathcal{E}$ and $\mathcal{F}$ in Section~\ref{sec:deformation_space}, we can use any neural network architecture that processes point clouds and produces a global shape feature and per-point features, respectively. Note that every \emph{three} rows of $\mathbf{A}_x \in \mathbb{R}^{3n \times k}$ determine the offset of each point \emph{independently} and thus can be predicted as a per-point feature. In our experiments (Section~\ref{sec:results}), we use PointNet~\cite{PointNet}; its classification architecture for the encoder $\mathcal{E}$
and the segmentation architecture for the deformation dictionary predictor $\mathcal{F}$. Figure~\ref{fig:pipeline} shows the entire pipeline.
A Siamese structure of the encoder $\mathcal{E}$ takes a pair of
source $x \in \overline{X}$ and target $y \in \overline{X}$ shapes
as input and predicts the deformation vector $\left( \mathcal{E}(y) - \mathcal{E}(x) \right) \in \mathbb{R}^k$. The dictionary predictor $\mathcal{F}$ takes \emph{only} the source shape as input and predicts the linear deformation dictionary $\mathcal{F}(x) \in \mathbb{R}^{3n \times k}$. For faster convergence in training, we normalize each column in the dictionary $\mathcal{F}(x)$ to have unit norm. The deformation of the source shape fitting to the target is predicted by computing point offsets $\left( \mathcal{F}(x) \left( \mathcal{E}(y) - \mathcal{E}(x) \right) \right) \in \mathbb{R}^{3n}$ and adding it to the source shape as described in Equation~\ref{eq:deformation}.
Additionally, if deformation handles are provided for each shape, the deformed source shape can also be projected to their space as described in Equation~\ref{eq:deformation_proj}. The fitting error from the output deformed source shape $d(x \rightarrow y)$ to the target shape $y$ is measured using Chamfer Distance (cf.,~\cite{Fan:2017, Achlioptas:2018, CycleConsistency,3DN,NeuralCages}):
%
\begin{align}
    \mathcal{L}_F := \text{Ch} \left( d(x \rightarrow y), y \right).
    \label{eq:fitting_loss}
\end{align}

We also follow the idea of Wang~\etal~\shortcite{3DN} and Yifan~\etal~\cite{NeuralCages} to preserve \emph{symmetries} of  man-made objects in the deformation. When a global reflection symmetry axis is given for the source shape, we flip the output deformed shape along the axis and minimize Chamfer distance to this:
\begin{align}
    \mathcal{L}_R := \text{Ch} \left( d(x \rightarrow y), \mathbf{R} \left( d(x \rightarrow y) \right) \right),
\end{align}
where $\mathbf{R}$ is the mirroring operation along the given reflection axis.

\rev{Additionally, we also support sparsity regularization losses to enforce structure in the output dictionaries. For example, in shape editing scenarios, the user may want to identify the essential correlations among the given deformation handles. To discover them, we can apply $l1$-loss to the output dictionary matrices \emph{after} projecting them to the given deformation handle space (as in Equation~\ref{eq:deformation_proj}):
\begin{align}
    \mathcal{L}_{S1} := \frac{1}{k} \|\mathbf{B}_x^\dagger \mathcal{F}(x)\|_1\,.
    \label{eq:l1_sparsity}
\end{align}
This loss can encourage the columns in the projected dictionary matrices to be sparse, so that they can capture strongly correlated deformation handles. Also, while we preset the number of columns in the dictionary $k$ during training (\ie,  the dimension of the latent space), we can find the minimal set of the columns necessary to handle all possible deformations by imposing a column-wise $l2,1$-loss to the projected dictionaries~\cite{Ding:2006,Nie:2010}:
\begin{align}
    \mathcal{L}_{S2,1} := \frac{1}{k} \|\mathbf{B}_x^\dagger \mathcal{F}(x)\|_{2,1},
    \label{eq:l21_sparsity}
\end{align}
which makes the norm of each column to be close to zero~\footnote{Note that we normalize the dictionary columns \emph{before} projection, and thus the norm of dictionary columns \emph{after} the projection can still be minimized (but cannot be zero).}. Empirically, we found that this loss also plays the role of $l1$-loss, sparsifying each column, even while making the training more stable. Hence, we only employ this $l2,1$-loss in our final training loss, which is defined as follows:

%
\begin{align}
    \mathcal{L} = \mathcal{L}_F + \mathcal{L}_R + w_{S2,1} \mathcal{L}_{S2,1},
\end{align}
where $w_{S2,1}$ is a weight for the sparsity loss. In Section~\ref{sec:qualitative_evaluations}, we analyze the effect of the sparsity loss.
}
%
%

\revfinal{Please note that we do not use any \emph{cycle-consistency} loss since the consistency of the dictionaries $\mathcal{F}(x)$ across the shapes automatically emerges during network training, as discussed in Section~\ref{sec:deformation_space}.}
In Section~\ref{sec:quantitative_evaluations_adriana}, we empirically evaluate cycle-consistency with a dataset where, as ground-truth, the point correspondences are known across the shapes. The experiment demonstrates that our method, without any loss function for consistency, performs even better results compared with the network of Groueix~\etal~\shortcite{CycleConsistency}, which explicitly optimizes using cycle-consistency losses.

\subsection{Comparison with Other Neural Deformations}
\label{sec:network_comparisons}

Recent neural networks learning target-driven deformations, such as 3DN~\cite{3DN}, Cycle Consistency~\cite{CycleConsistency}, and Neural Cages~\cite{NeuralCages}, have an architecture analogous with ours in that they take a source-target pair of shapes as input, compute a latent vector for the shape difference, and apply it to the source shape. We discuss the main differences between these methods and ours, and describe how the differences affect performance and usage in downstream applications.

\paragraph{Learning Shape-Dependent Variation Space}
While both our network and the others are trained to learn deformation from one shape to the other, we not only learn how to fit the input to the target but also discover the plausible variation space of the input shape, which is represented as a linear space with the deformation dictionary $\mathbf{A}_x$. Hence, in shape editing, when the deformation handles are given as linear deformation functions, the user's input can be easily projected back to the learned variation space as described in Section~\ref{sec:projection} and also demonstrated in Section~\ref{sec:qualitative_evaluations}. This is an additional capability of our method compared to the other methods. \rev{Moreover, our deformation dictionary $\mathbf{A}_x$ captures strong correlations among the given deformation handles with additional regularization loss (see Equation~\ref{eq:l21_sparsity}), and thus provides more intuitive interpretation of the learned deformation space, as discussed in Section~\ref{sec:qualitative_evaluations}.}

\paragraph{Factorizing Deformation Representation}
We also found that the factorization of shape offset representation into a source-dependent dictionary $\mathcal{F}(x)$ and a latent vector for a pair of shapes $\left( \mathcal{E}(y) - \mathcal{E}(x) \right)$ gives a better performance in the fitting. See Section~\ref{sec:quantitative_evaluations} for quantitative evaluations. While Neural Cages~\cite{NeuralCages} have a similar factorization to ours, the others (3DN~\cite{3DN} and Cycle Consistency~\cite{CycleConsistency}) immediately combine the information of the pair to the source shape in the network without separately manipulating the source shape beforehand.

\paragraph{Enforcing Affine Space Properties}
To attain the affine space properties in the latent space, in our framework, the deformation from one shape to the other is represented as the \emph{subtraction} of two latent codes $\left( \mathcal{E}(y) - \mathcal{E}(x) \right)$, as described in Section~\ref{sec:deformation_space}. In all the other methods, however, the deformation is learned by first \emph{concatenating} two latent codes and processing it in the next layers. In Section~\ref{sec:quantitative_evaluations}, we conduct an ablation study, changing the computation of the latent deformation vector in our framework, similarly with the other methods. The results demonstrate that our architecture enforcing the affine space properties produces more plausible shapes in the deformation transfer compared with the other methods and the architecture in the ablation study.

%% file: sections/results.tex

\section{Results}
\label{sec:results}

We utilize our framework in the context of shape \emph{co-editing} application. Also, we quantitatively evaluate our method and compare against state-of-the-art shape deformation techniques.

\subsection{Qualitative Evaluations}
\label{sec:qualitative_evaluations}

\begin{figure}[t!]
\centering
\setlength{\tabcolsep}{0em}
\def\arraystretch{0.0}
\newcolumntype{Y}{>{\centering\arraybackslash}X}
{\small
\begin{tabularx}{\columnwidth}{YYYYY}
  \shortstack{Airplane\\(2390)} &
  \shortstack{Car\\(1200)} &
  \shortstack{Chair\\(1927)} &
  \shortstack{Sofa\\(947)} &
  \shortstack{Table\\(1857)} \\
  \multicolumn{5}{c}{\includegraphics[width=\columnwidth]{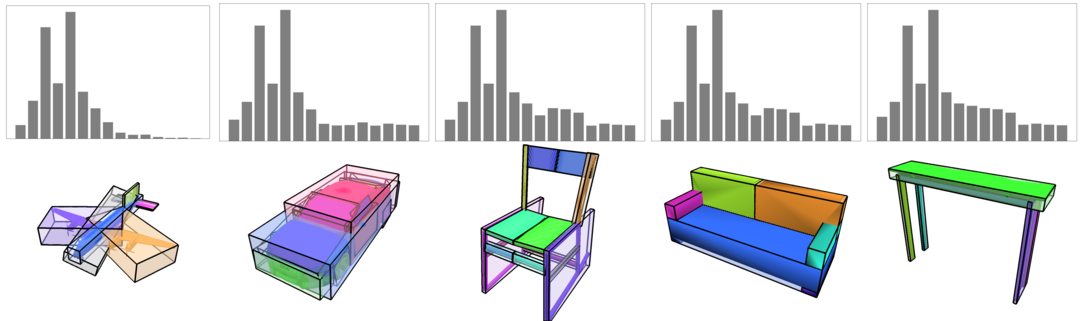}} \\
\end{tabularx}
}
\vspace{-0.5\baselineskip}
\caption{ShapeNet part bounding box dataset. Each column shows the number of shapes, a histogram of the number of components, and a sample object with part bounding boxes for each category. We preprocess raw CAD models as described in Section~\ref{sec:qualitative_evaluations} and collect models with number of components is in the range of $[2, 16]$.}
\label{fig:data_stats}
\end{figure}

\begin{figure*}[t!]
\centering
\includegraphics[width=\textwidth]{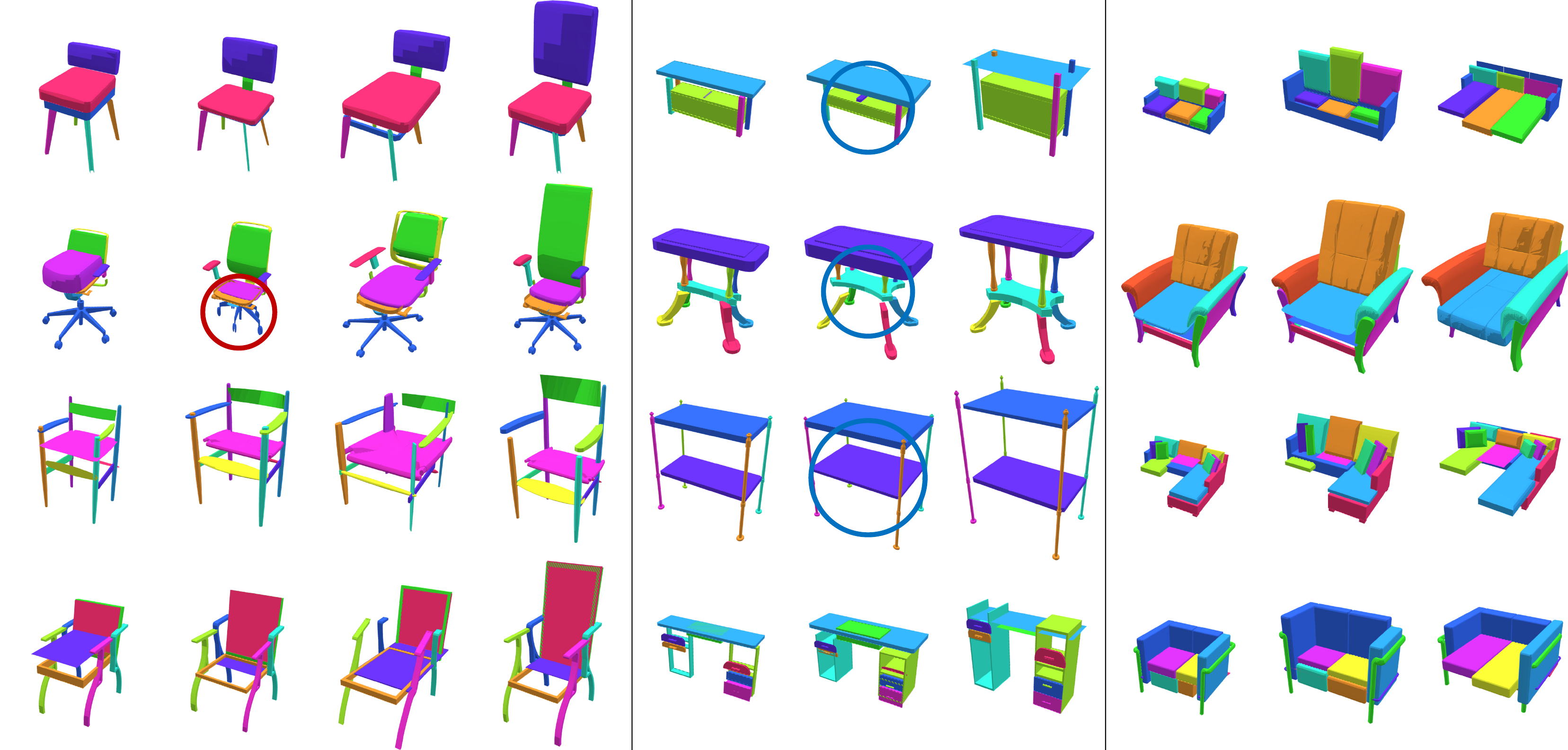}
\vspace{-\baselineskip}
\caption{\rev{Elements in the learned deformation dictionary. The columns are deformations along some elements in the dictionary (columns of $\mathbf{A}_x$). Each element shows a natural \emph{mode} of shape variations changing local parts, and the variations of each element are also \emph{consistent} across the shapes despite the difference in styles of the shapes. The second column of chairs shows scaling of swivel leg (red circle), and the element does not affect the shapes not including the swivel leg. Also, the second column of tables lifts up the shelf in the middle (blue circle) and makes no change if the part does not exist. Refer to the supplemental video for animated visualizations.}}
\label{fig:dictionary}
\end{figure*}

\paragraph{Dataset and Network Training}
We experiment with five categories from the ShapeNet repository~\cite{ShapeNet}, namely Airplane, Car, Chair, Sofa, and Table.
Most of the 3D models in ShapeNet are composed of smaller parts that have simpler geometry. Thus, we compute oriented bounding boxes for parts using the Trimesh library~[\citeauthor{trimesh}] and take \emph{translation} and anisotropic \emph{scales} along each local coordinate as our deformation handles. Figure~\ref{fig:data_stats} shows the number of shapes in each category (in parentheses), the distribution of the number of components in each shape (first row), and a sample object with part bounding boxes (second row). Before computing the bounding boxes, we run a preprocessing to merge small components to their closest neighbors and also combine overlapping components, as illustrated in the work of Sung~\etal~\shortcite{Sung:2017}. However, we do not group symmetric parts in order to have more freedom in editing each part.
(Symmetries will be \emph{learned} by our networks, as illustrated in Section~\ref{sec:qualitative_evaluations} and ~\ref{sec:structure_discovery}.)
After preprocessing, we take models with a number of component parts in the range of $[2, 16]$. We do not normalize the scale of each shape individually to see natural shape variations (e.g., making a chair to a bench without decreasing the height), but all shapes are in the range of $[-0.5, 0.5]$ in each axis.

We train the networks for each category.
We sample $2K$ points randomly over each shape and feed the networks with random pairs of the source and target shapes. Training, validation, and test sets are randomly split with the 85-5-10 ratio. We set the dimension of latent space (i.e., the number of dictionary elements) $k$ to \rev{$64$ and the weight of sparsity regularization loss $w_{S2,1}$ to $10$.}

Since we aim to deal with \emph{uncurated} datasets where no manual annotations are present, we do \textit{not} use any part labels in ShapeNet (unlike those used by Mo~\etal~\shortcite{Mo:2019:StructEdit}) or any correspondence supervision in network training. Also, such part labels are often insufficient to infer correspondences across the deformation handles.

\paragraph{Deformation Dictionaries}
\rev{We illustrate the learned deformation dictionary $\mathbf{A}_x$ in Figure~\ref{fig:dictionary}. Each column visualizes shapes deformed along some elements in the dictionaries. Thanks to the sparsity regularization loss (Equation~\ref{eq:l21_sparsity}), the elements show local and natural deformation \emph{modes} and correlations among the given deformation handles. For example, the first and third columns of chairs translate and scale the seat along up and front directions, respectively, while preserving connectivities with the other parts. The last column also elongates the back part upward. Similar local deformations are also observed in sofas. Interestingly, the second column scales only swivel leg (red circle), and it does not affect the shape with a different type of leg. Similarly, in the second column of tables, the element translates the shelf in the middle along the up direction and does not make any changes if the shape does not include the shelf.
The supplementary video shows how the shapes vary along the dictionary elements in animations.}

\paragraph{Shape Co-Editing}

\begin{figure*}[t!]
\centering
\setlength{\tabcolsep}{0em}
\def\arraystretch{0.0}
\newcolumntype{Y}{>{\centering\arraybackslash}X}
{\small
\begin{tabularx}{\textwidth}{YYYYYYY}
  Source & \shortstack{Edited\\(w/ boxes)} & \shortstack{Projected\\(w/ boxes)} & Projected & New Source & \shortstack{Transferred\\(w/ boxes)} & Transferred \\
  \midrule
  \multicolumn{7}{c}{\includegraphics[width=\textwidth]{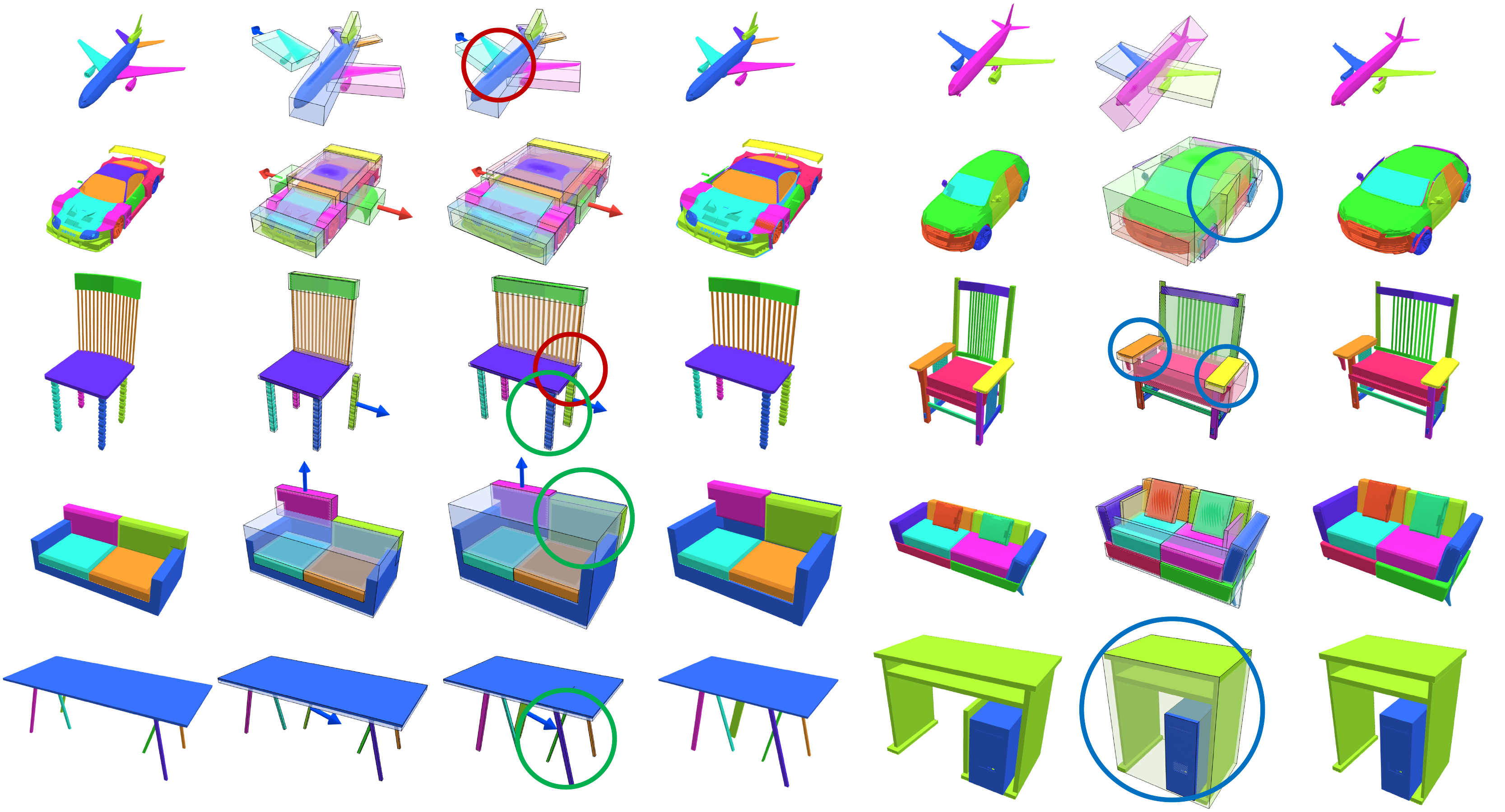}} \\
\end{tabularx}
}
\vspace{-0.5\baselineskip}
\caption{Shape editing projection and transfer results. The first column is the source shape, the second column is the user edit (blue and red arrows mean translation and scaling along the direction, respectively), the third and fourth columns are the results of projection (with and without the part bounding boxes), the fifth column is the new source shape, and the sixth and seventh columns are the results of transferring the projected deformation to the new source shape (with and without the part bounding boxes). The projection adjusts the rest of the deformation handles while preserving part connectivity (red circles) and symmetry (green circles). Also, the projected deformations are naturally transferred to the new shape despite the different part structures  (blue circles). See the supplemental video for more examples.}
\label{fig:project_and_transfer}
\end{figure*}

We also describe how our framework can be employed for interactive shape co-editing and demonstrate qualitative results. We consider a scenario that the user modifies a source shape through one of the given deformations handles at each step. Given the user input, the system (i)~first automatically \emph{snaps} the edited shape to the plausible shape space while fixing the modified parameter as described in Section~\ref{sec:projection}. When solving the least square in Equation~\ref{eq:projection_lsq}, we use a constraint that all scale parameters should be greater than zero. Then, (ii)~the projected deformation is  transferred to the desired new source shape in the same category.

Figure~\ref{fig:teaser} and~\ref{fig:project_and_transfer} show some examples of the deformation projection and transfer. In Figure~\ref{fig:project_and_transfer}, given a source shape (first column), we randomly select one of the deformation handles and perturb the parameter (second column). If the selected handle is a translation along one of the local coordinates of a box (blue arrow in the figure), we vary the parameter in the range of [$-0.5+t$, $0.5+t$], where $t$ is the given number. If the select handle is a scaling (red arrow in the figure), we pick a random number in the range of [$0.5s$, $1.5s$], where $s$ is the default scale. Then, the modified shape is projected to the learned plausible shape space (third and fourth column). Given another source shape (fifth column), the projected deformation is transferred (sixth and seventh columns) by taking the latent deformation vector and applying it to the new source shape with the learned shape-dependent action. The results show that the projection adjusts the rest of the parameters in a way to maintain the shape structure such as symmetry and part connectivity. For instance, right-wing of the airplane (first row) and a leg of the chair (third row) and the table (last row) are properly attached to the other parts after the projection while preserving symmetry. Also, the deformation on the source shape is naturally transferred to the new source shape even when the geometry or the deformation handles are different.
For example, the tables in the last row have different part structure, but the translation of legs is naturally adopted to the new shape as decreasing the width.
Refer to the supplemental video for more examples.

\rev{
\paragraph{Effect of Sparsity Regularization Losses}
\begin{figure*}[t!]
\centering
\begin{minipage}[t]{.33\textwidth}
  \centering
  \includegraphics[width=\textwidth]{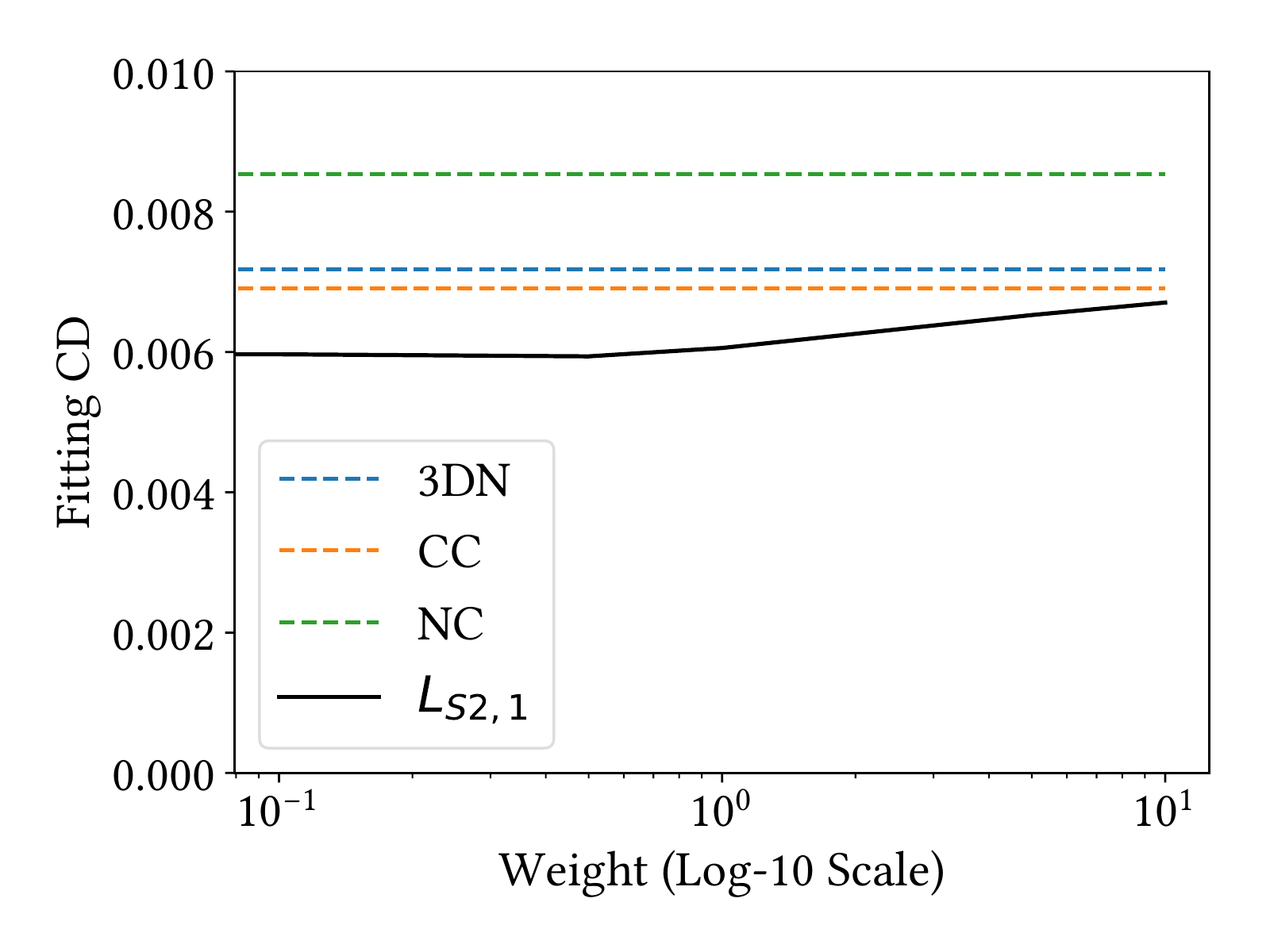}
  \vspace{-2.5\baselineskip}
  \caption*{(a)}
\end{minipage}\hfill%
\begin{minipage}[t]{.33\textwidth}
  \centering
  \includegraphics[width=\linewidth]{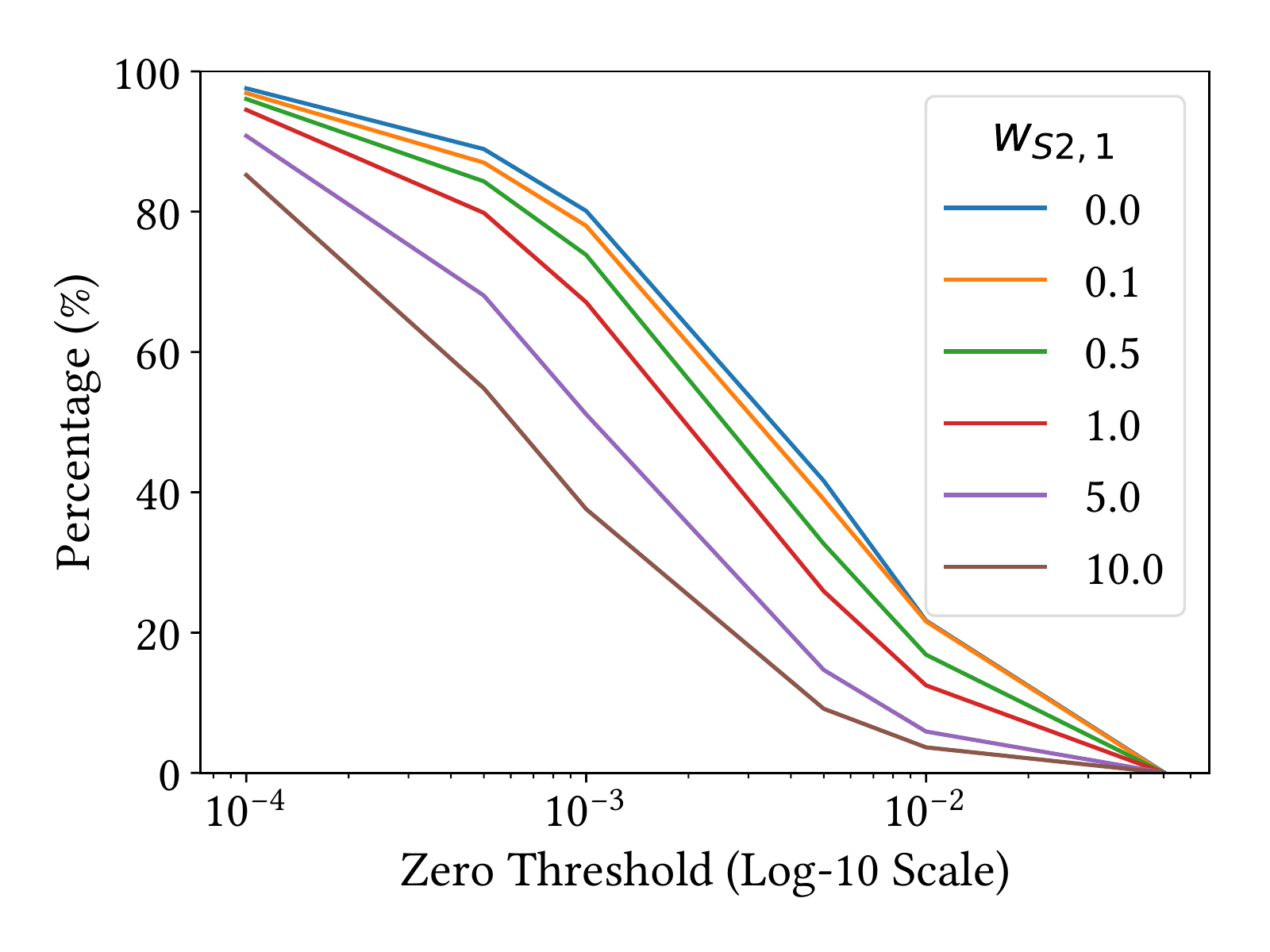}
  \vspace{-2.5\baselineskip}
  \caption*{(b)}
  \label{fig:s3dis_confusion_matrix}
\end{minipage}\hfill%
\begin{minipage}[t]{.33\textwidth}
  \centering
  \includegraphics[width=\linewidth]{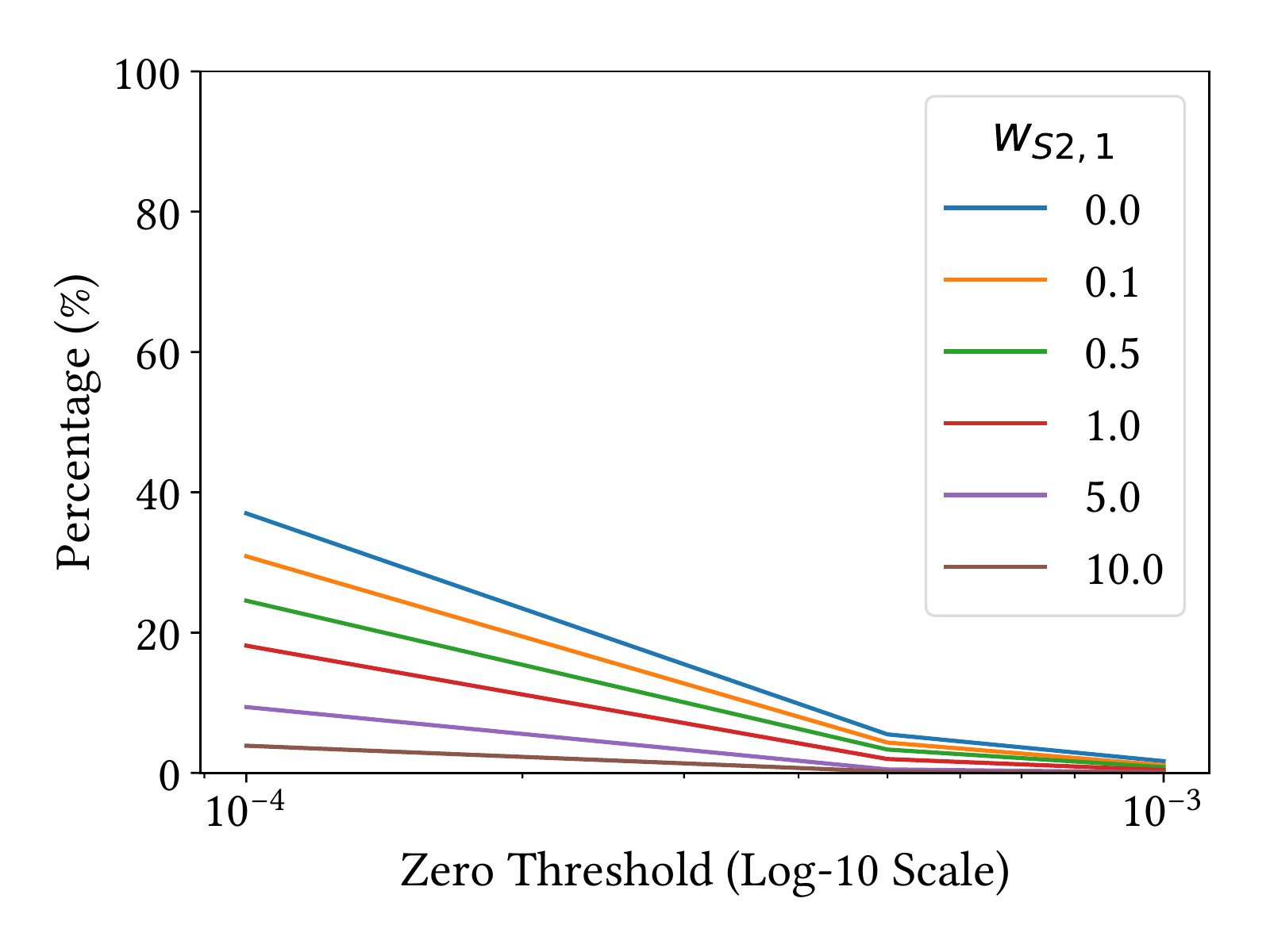}
  \vspace{-2.5\baselineskip}
  \caption*{(c)}
\end{minipage}
\vspace{-0.5\baselineskip}
\caption{\rev{Effect of Sparsity Regularization Losses. (a) Fitting error comparison when varying the weight of sparsity regularization loss in Equation~\ref{eq:l21_sparsity}. The x-axis is the weight (in log-10 scale), and the y-axis is the fitting error measured as Chamfer distance. The dot lines indicate the fitting errors of baseline methods: 3DN~\cite{3DN}, Cycle Consistency (CC)~\cite{CycleConsistency} and Neural Cages (NC)~\cite{NeuralCages}. (b) The percentage of non-zero \emph{elements} with varying weights. The x-axis is the zero threshold (in log-10 scale). (c) The number of non-zero \emph{columns} with varying weights.}}
\label{fig:sparsity_loss}
\end{figure*}
We analyze the effect of sparsity regularization loss in Equation~\ref{eq:l21_sparsity} by varying its weight. Figure~\ref{fig:sparsity_loss} (a) shows the mean fitting error (Chamfer distance) of $1k$ random chair pairs when varying the weight from $0.1$ to $10$. While the fitting error increases with a larger weight, the errors with large weights are smaller than the ones of baseline deformation methods (see Section~\ref{sec:quantitative_evaluations}). In Figure~\ref{fig:sparsity_loss} (b) and (c), we show how many non-zero elements and columns we obtain in the projected dictionaries $\mathbf{B}_x^\dagger \mathcal{F}(x)$ when increasing the weight and also varying the threshold of zero. The numbers of non-zero elements and columns dramatically decrease, indicating that the network discovers more strongly correlated deformation handles, while still learning all possible deformations.
}

\subsection{Quantitative Evaluations Using ShapeNet}
\label{sec:quantitative_evaluations}
Next, we quantitatively evaluate our method using ShapeNet~\cite{ShapeNet} dataset and compare with the other baseline methods.

\paragraph{Dataset and Network Training}
For the quantitative evaluations, we use the same preprocessed dataset of ShapeNet with Groueix~\etal~\shortcite{CycleConsistency}. This dataset contains five categories: Airplane, Car, Chair, Lamp, and Table. The difference of this dataset with the one used in Section~\ref{sec:qualitative_evaluations} is that it normalizes each point cloud to be fit in a uniform cube, which size is $2$ for each axis. All networks including ours and baselines are trained as described in Section~\ref{sec:qualitative_evaluations}, by taking $2k$ random sample points in each shape and feeding random source-target shape pairs in each category. To maximize the fitting capability, in this experiment, we set the dimension of the latent space in our framework, $k$, to 512 \rev{and disable sparsity regularization loss ($w_{S2,1}=0$).}
Also, in our framework, we do not use the projection to the deformation handle space described in Equation~\ref{eq:deformation_proj} during the network training, and the effect of the projection is analyzed in Section~\ref{sec:effect_of_projection}.

\paragraph{Baselines}
We compare our method with non-rigid ICP by Huang~\etal~\shortcite{Huang:2017} and three recent neural-network-based deformation methods mentioned in Section~\ref{sec:network_comparisons}: 3DN~\cite{3DN}, Cycle Consistency (CC)~\cite{CycleConsistency}, and Neural Cages (NC)~\cite{NeuralCages}. We take point clouds as input to all networks and do not use the differentiable mesh sampling operator introduced in 3DN~\cite{3DN}, which can be easily plugged into any of the networks.
All the baseline neural-net-based methods have a part in their architecture to compute deformation (\ie shape difference) from the given source and target shapes and apply it to the source shape. Thus, we implement the deformation transfer of the other methods as applying the given source-target pair information to the other source shape, similarly with our method.

\paragraph{Ablation Study}
We test the impact of enforcing affine space properties by modifying the part of computing latent deformation vector in our networks. As mentioned in Section~\ref{sec:network_comparisons}, instead of taking the \emph{difference} of two latent codes $\left( \mathcal{E}(y) - \mathcal{E}(x) \right)$, we concatenate latent codes of source and target shapes and process it to produce the deformation vector~\footnote{Within the PointNet~\cite{PointNet} classification architecture, we concatenate global features of the shapes produced from the max-pooling layer and then pass it through the same next MLP layers. The final output replaces $\left( \mathcal{E}(y) - \mathcal{E}(x) \right)$ in our original architecture.}.

\paragraph{Target-Driven Deformation and Transfer}

\begin{figure*}[t!]
\centering
\setlength{\tabcolsep}{0em}
\newcolumntype{Y}{>{\centering\arraybackslash}X}
{\footnotesize
\begin{tabularx}{\textwidth}{Y|Y|YYYYY|Y|YYYY}
  \multirow{2}{*}{Source} &
  \multirow{2}{*}{Target} &
  \multicolumn{5}{c|}{Deformed} &
  \multirow{2}{*}{New Source} &
  \multicolumn{4}{c}{Transferred} \\
  \cline{3-7}\cline{9-12}
  & & NR-ICP & 3DN & CC & NC & \DSN & & 3DN & CC & NC & \DSN \\
  \midrule
  \multicolumn{12}{c}{\includegraphics[width=\textwidth]{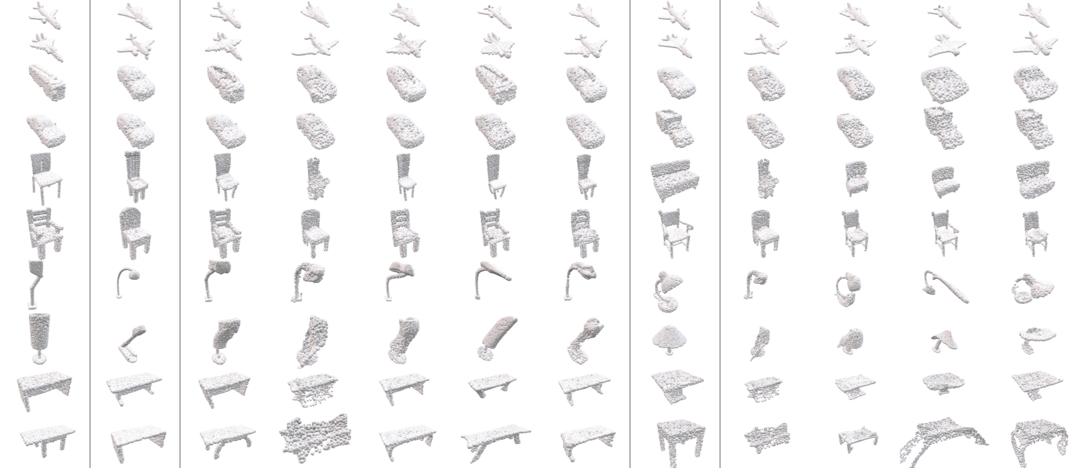}} \\
\end{tabularx}
}
\vspace{-0.5\baselineskip}
\caption{Examples of qualitative evaluation results. The first and second columns are the source and target shapes, the next five columns are deformed source shapes fitted to the target, resulted by different methods, the next eighth column is the new source shape, and the next four columns are the results of deformation transfer from the source-target pair to the new source. The baseline methods compared with our \DeformSyncNet~(\DSN) are Non-rigid ICP (NR-ICP)~\cite{Huang:2017}, 3DN~\cite{3DN}, Cycle Consistency (CC)~\cite{CycleConsistency} and Neural Cages (NC)~\cite{NeuralCages}.}
\label{fig:deform_and_transfer_pc}
\end{figure*}

\input{tables/deform_and_transfer_pc.tex}

We compare the methods in two aspects: fitting accuracy of deformations and plausibility of deformation transfer results.

For the fitting accuracy, we perform target-driven deformation, deforming a \emph{source} shape to fit it to a \emph{target} shape, with random $1,000$ pairs of source and target test shapes in each category. Then, we measure the average Chamfer distance between the target and the deformed source shape (Fitting CD). Also, inspired by Groueix~\etal~\shortcite{CycleConsistency}, we evaluate the fittings by finding the closest points from the deformed source shape to the target and measuring mean Intersection over Union (mIoU) of semantic parts provided by Yi~\etal~\shortcite{Yi:2016}. This metric demonstrates how beneficial each method at unsupervised co-segmentation or few-shot segmentation. 

For the plausibility of deformation transfer results, the quantitative evaluation is challenging since it is not precisely determined without involving human perception. In our evaluation, we follow the ideas of Achlioptas~\etal~\shortcite{Achlioptas:2018} evaluating generative models. Achlioptas~\etal introduce two \emph{data-driven} evaluation metrics: Minimal Matching Distance (MMD-CD) and Coverage (Cov-CD). Minimal Matching Distance is the average of the Chamfer distance from each generated shape to its the closest shape in the reference dataset, indicating how likely the output shape looks like a real shape. Coverage is the proportion of the shapes in the reference dataset that are closest from each generated shape, showing how many variations are covered by the generated shapes. For these two, we randomly choose $200$ pairs of source and new source shapes, and for each of the $10$ target shapes, we transfer the source-to-target deformation to the new source shape. Then, we measure the metrics by taking the training set as the reference dataset; the averages for all target shapes are reported.

We report all results in Table~\ref{tbl:deform_and_transfer_pc}. Bold is the best result, and underscore is the second-best result. Note that our \DeformSyncNet~is the only method that shows outstanding performance both in the fitting accuracy and in the plausibility of deformation transfer results. For the Fitting CD and mIoU, \DeformSyncNet~gives better accuracy in most of the categories compared with the other methods. The network in the ablation study, concatenating two latent codes, is the only one that shows the better fitting accuracy than \DeformSyncNet~in terms of Fitting CD, but its performances in other metrics are poor, particularly in MMD-CD and Cov-CD. For the MMD-CD and Cov-CD, our \DeformSyncNet~also outperforms the other methods. The only competitor is Neural Cages~\cite{NeuralCages}, but it gives inferior accuracy of the fitting, as shown in Fitting CD.

Figure~\ref{fig:deform_and_transfer_pc} illustrates some results of quantitative evaluation. Our \DeformSyncNet~shows the best fitting results in most of the cases and also captures well the difference between the source and target shapes in the transfer. For instance, the source and target chairs in the fifth row have differences in the width and height of the back part, and these differences are correctly transferred to the new shape, which has a different global structure. Also, in the sixth row, ours combines chair arms of the source shape to seat in the deformation, and this change is properly transferred to the new shape. The other methods tend not to transfer the \emph{difference} but to do target-oriented deformation, making the new source shape close to the target shape.

\bigbreak
\paragraph{Parallelogram Consistency Test}
\label{sec:parallelogram}
\begin{figure*}[ht!]
\centering
\setlength{\tabcolsep}{0em}
\newcolumntype{Y}{>{\centering\arraybackslash}X}
{\footnotesize
\begin{tabularx}{\textwidth}{Y|Y|Y|YY|YY|YY|YY}
  \multirow{2}{*}{$x$} &
  \multirow{2}{*}{$y$} &
  \multirow{2}{*}{$z$} &
  \multicolumn{2}{c|}{3DN~\cite{3DN}} &
  \multicolumn{2}{c|}{CC~\cite{CycleConsistency}} &
  \multicolumn{2}{c|}{NC~\cite{NeuralCages}} &
  \multicolumn{2}{c}{\DeformSyncNet} \\
  \cline{4-11}
  & & &
  $z \oplus \overrightarrow{x y}$ & $y \oplus \overrightarrow{x z}$ & 
  $z \oplus \overrightarrow{x y}$ & $y \oplus \overrightarrow{x z}$ & 
  $z \oplus \overrightarrow{x y}$ & $y \oplus \overrightarrow{x z}$ &
  $z \oplus \overrightarrow{x y}$ & $y \oplus \overrightarrow{x z}$ \\
  \midrule
  \multicolumn{11}{c}{\includegraphics[width=\textwidth]{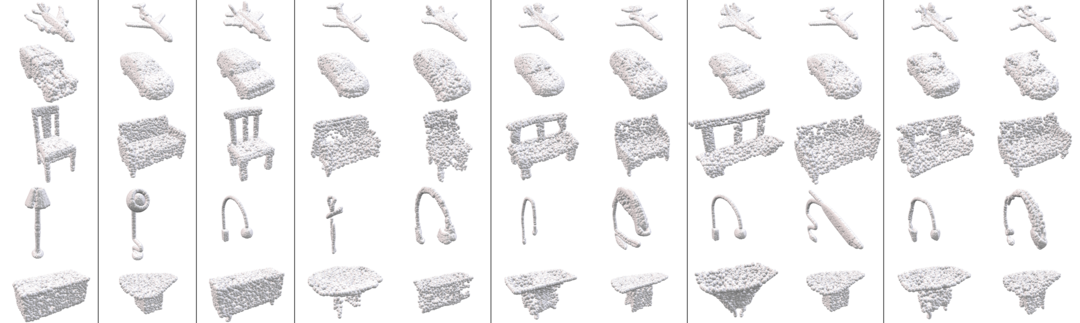}} \\
\end{tabularx}
}
\vspace{-0.5\baselineskip}
\caption{Qualitative results of parallelogram consistency test. Our \DeformSyncNet~provides the most consistent result of deformation given two different pathways: $z \oplus \protect\overrightarrow{x y}$ and $y \oplus \protect\overrightarrow{x z}$. See Section~\ref{sec:parallelogram} for the details.}
\label{fig:parallelogram}
\end{figure*}

\setlength{\columnsep}{5pt}
\begin{wrapfigure}{r}{0.25\columnwidth}
  \vspace{-\baselineskip}
  \begin{center}
    \includegraphics[width=0.25\columnwidth]{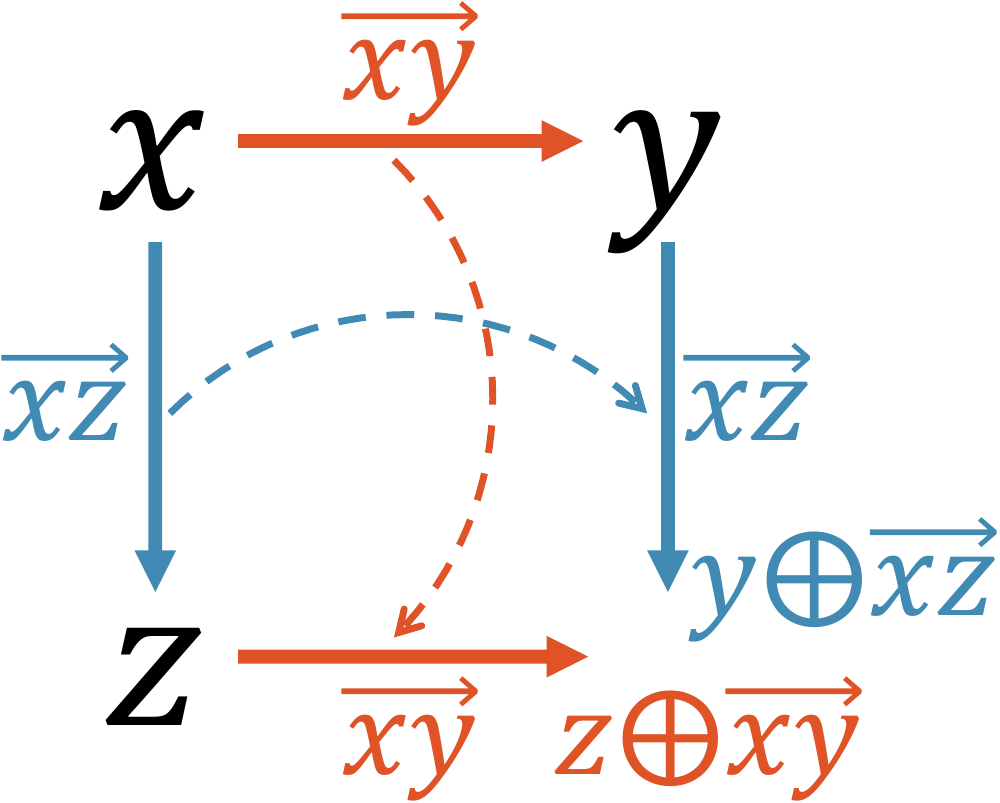}
  \end{center}
  \vspace{-\baselineskip}
\end{wrapfigure}
One of the beneficial properties of affine space is the \emph{path invariance}, producing the same deformation regardless of the pathway from the starting point to the destination. This property is desirable in the downstream applications since it allows the user to explore the latent shape variation space freely without specifying the order of steps.
We test this property by taking a random \emph{parallelogram} in the latent space. A triplet of shapes $x$, $y$, and $z$ are given, and deformations are transferred in two ways: $\overrightarrow{x y}$ to $z$ and $\overrightarrow{x z}$ to $y$ (see inset). Then, we verify whether these two results are the same by measuring Chamfer distance.

The quantitative results comparing with other methods are reported in Table~\ref{tbl:parallelogram}. Our \DeformSyncNet~gives much smaller differences between two pathways of deformations compared with the other methods. Cycle Consistency~\cite{CycleConsistency} using regularization losses for the consistency provides the second-best results in all categories, but its fitting distance is much larger than ours. Figure~\ref{fig:parallelogram} illustrates some of the qualitative results.

\subsection{Quantitative Evaluations and User Study Using Parametric 3D Models}
\label{sec:quantitative_evaluations_adriana}
For further quantitative analysis and user study, we experiment with parametric 3D models provided by Schulz~\etal~\shortcite{Schulz:2017} and compare our method with the other methods.

\paragraph{Dataset and Network Training}
The dataset of Schulz~\etal~\shortcite{Schulz:2017} contains 74 parametric models, each of which has its own deformation parametrization. For each model, we uniformly sample $2k$ points and generate $1k$ variants with random parameters; $128$ shapes out of them are used as the test set. The main difference of this dataset with ShapeNet in Section~\ref{sec:quantitative_evaluations} is that point correspondences are \emph{known} across the shapes since they are generated from the same parametric models. The point correspondences are \emph{not used} as supervision in any experiment, but will be used as ground-truth in evaluations. During training, we do \emph{not} project the deformed shape to the given deformation parameter space (see Equation~\ref{eq:deformation_proj}).

\input{tables/parallelogram.tex}

\begin{figure}[t!]
\centering
\vspace{-0.5\baselineskip}
\begin{minipage}[t]{.5\columnwidth}
\includegraphics[width=\linewidth]{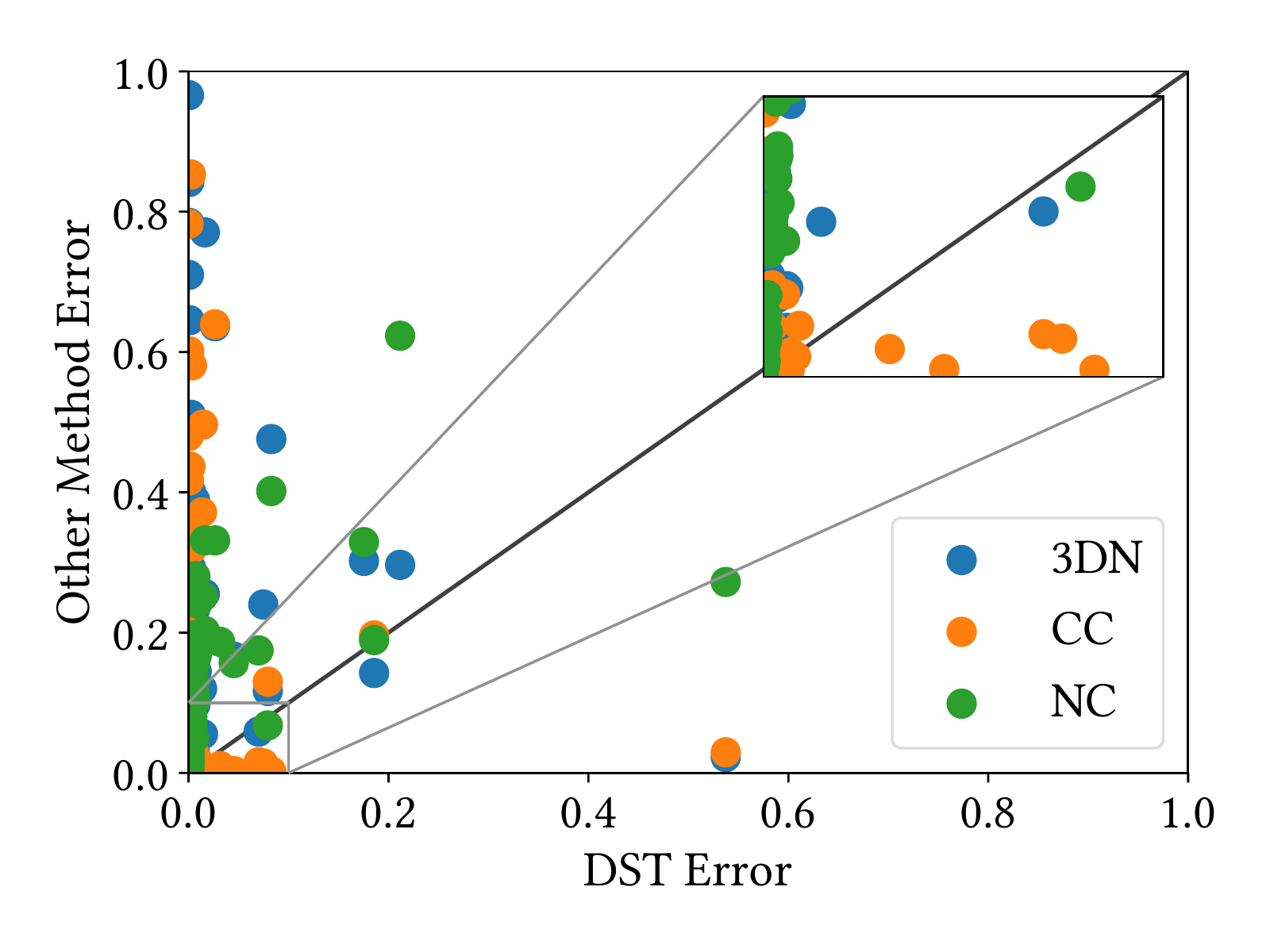}
\vspace{-2.5\baselineskip}
\caption*{(a)}
\end{minipage}\hfill
\begin{minipage}[t]{.5\columnwidth}
\includegraphics[width=\linewidth]{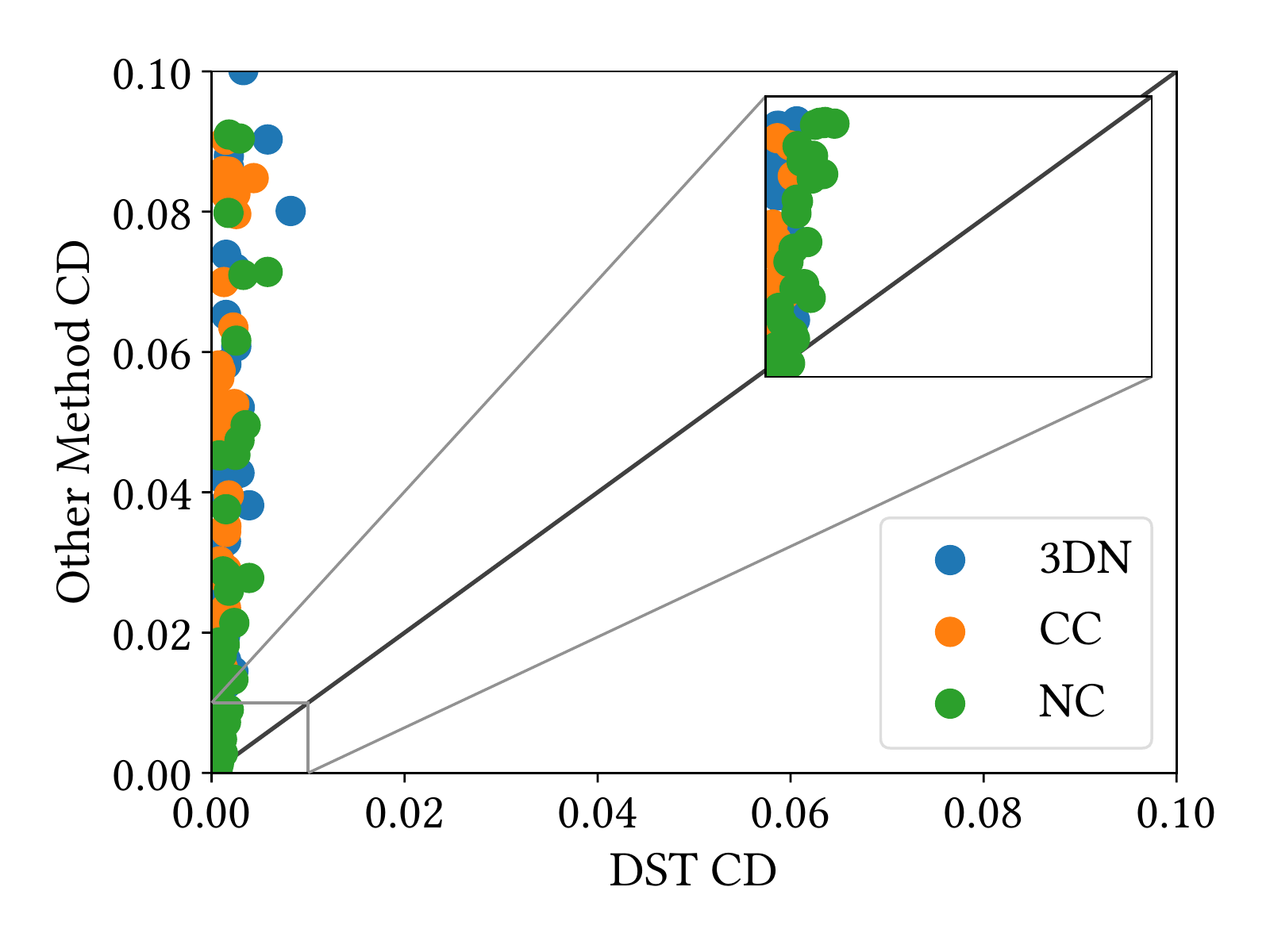}
\vspace{-2.5\baselineskip}
\caption*{(b)}
\end{minipage}
\vspace{-\baselineskip}
\caption{\rev{Two-way consistency and deformation transfer results using parametric models from Schulz~\etal~\shortcite{Schulz:2017}. (a) Comparison of two-way cycle consistency errors between \DeformSyncNet~and the other methods. Each dot is a result of one of 74 parametric models, and x- and y-axes are \DeformSyncNet~and the other methods, respectively. 
Note that in this visualization, above the diagonal line indicates that the corresponding method performs worse than ours. 
%
\revfinal{Compared with 3DN, Cycle Consistency (CC), and Neural Cages (NC), ours gives smaller errors in 70, 58, and 71 cases, respectively, out of total 74 cases. }
(b) Comparison of deformation transfer errors (Chamfer distance between the prediction and the ground truth).
Again, ours gives smaller errors in 74, 69, and 73 cases compared with 3DN, Cycle Consistency (CC), and Neural Cages (NC), respectively.}}
\label{fig:two_way_consistency_deformation_transfer}
\vspace{-\baselineskip}
\end{figure}

\paragraph{Two-Way Consistency Evaluation}
As discussed in Section~\ref{sec:network_design}, we empirically verify the effect of our network design enforcing consistency in the deformation, compared with Cycle Consistency in Groueix~\etal~\shortcite{CycleConsistency} leveraging dedicated loss functions for the consistency and also the other baseline methods. We train the networks for the shapes of \emph{each} parametric model and measure the two-way consistency error for a point cloud pair ($x$, $y$) as follows:
\begin{align*}
    \frac{1}{n} \sum_i \| \left( d_i(x \rightarrow y) - x_i \right) + \left( d_i(y \rightarrow x) - y_i \right) \|^2,
\end{align*}
where $x_i \in \mathbb{R}^3$ and $\left( d_i(x \rightarrow y) - x \right) \in \mathbb{R}^3$ are $i$-th point of the point cloud $x$ and its new position after deformation toward $y$, respectively, $n$ is the number of points, and here we assume that all point clouds are ordered consistently based on the given point correspondences. We computed the mean of this two-way consistency error for 50 pairs, which are the first 50 of the largest Chamfer distances among $1k$ randomly generated pairs.
Figure~\ref{fig:two_way_consistency_deformation_transfer} (a) shows a comparison between the results of our method and the baseline methods. Each dot indicates the consistency error for a single parametric model, and the x-axis and y-axis are for our \DeformSyncNet~and the other methods, respectively.
Our \DeformSyncNet~gives smaller two-way consistency errors than the other methods including Cycle Consistency in most cases: 70, 58, and 71 cases out of the 74 models compared with 3DN, Cycle Consistency, and Neural Cages, respectively. 

\paragraph{Quantitative Evaluation of Deformation Transfer}
Using the given deformation parametrization, we also quantitatively evaluate the performance of deformation transfer. If we choose all of the source, target, and destination (the new shapes to transfer deformation) shapes from the same parametric model, the ground truth of the deformation transfer can be computed by transferring the difference of parameters from source to target to the destination shape. Given the 50 test pairs per model above, we randomly pick another shape as the destination and compute Chamfer distance between the predicted and the ground truth shape. Figure~\ref{fig:two_way_consistency_deformation_transfer} (b) illustrates the mean Chamfer distance; x-axis is \DeformSyncNet,~and y-axis is the other methods. In most cases, \DeformSyncNet~gives a magnitude smaller distances compared with the other methods. The numbers of cases out of the 74 models when \DeformSyncNet~outperforms 3DN, Cycle Consistency, and Neural Cages are 74, 69, and 73, respectively.

\begin{figure}[t!]
\centering
\fbox{\includegraphics[width=0.9\columnwidth]{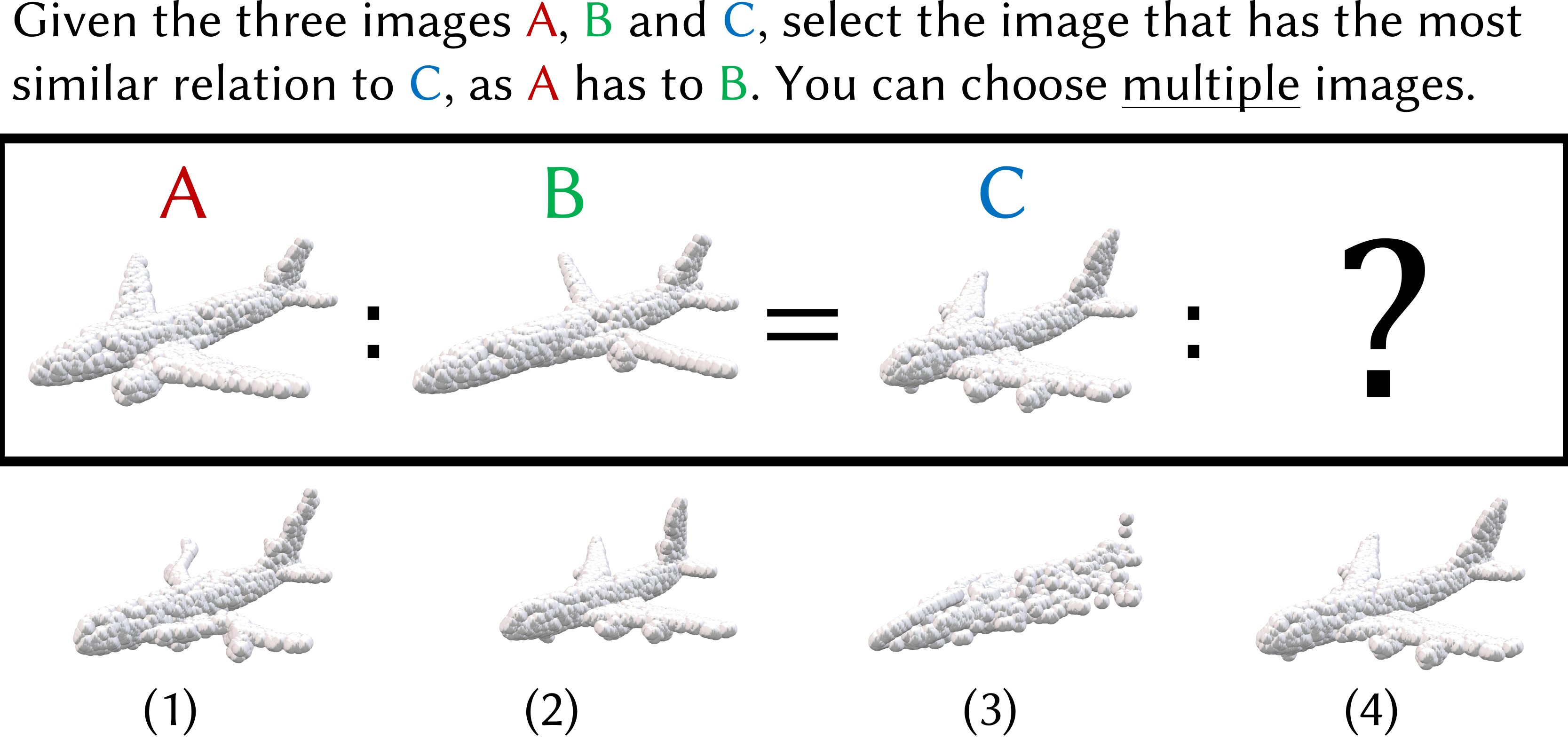}}
\vspace{-0.5\baselineskip}
\caption{\rev{User study example. The source (A) and target (B) shapes in the first row are sampled from the same parametric model, while the destination (C) shape is sampled from the other parametric model in the same category. The results of 3DN, Cycle Consistency, Neural Cages, and our \DeformSyncNet~are shown in the second row with a randomly permuted order. Multiple choices were allowed.}}
\label{fig:user_study_example}
\vspace{-\baselineskip}
\end{figure}

\paragraph{User Study for Deformation Transfer}
We further assess the quality of deformation transfer results by taking the destination shape from a \emph{different} parametric model (but in the same category) with the source and target shapes. Since we cannot compute the ground truth in this case, we conducted a user-study on Amazon Mechanical Turk.
Among the 74 parametric models, we grouped 8 Airplane, Chair, and Table models per category and trained all the networks for each group, while still taking the source and target shape from the same parametric model. But in the test time, the destination shape is randomly selected from the other model in the same group --- note that we take the source and target shapes from the same parametric model so that the participants can clearly identify the difference. From the same 50 test pairs per model above, in total 1,200 ($50 \times 8$ models $\times 3$ groups), we randomly select $10\%$ ($120$ pairs), assign the random destination shape, and create user study questions as shown in Figure~\ref{fig:user_study_example}. 
For each question, we ask human subjects (Turkers) to select \emph{at least one} among four shown objects: the outputs of 3DN, Cycle Consistency, Neural Cages, and our \DeformSyncNet.
The associations between the methods and the objects are hidden to the Turkers, and the order of the objects is randomized. The Turkers are also encouraged to choose \emph{multiple} objects if they think more than one options are equally good. In total, we collected 1,200 responses from a pool of 100 different participants (each question was answered by 10 distinct Turkers).

In the results, our \DeformSyncNet~was selected in $54.7\%$ of all 1,200 responses, whereas 3DN, Cycle Consistency, Neural Cages were selected in $30.9\%$, $4.0\%$, and $46.6\%$, respectively. (Note that we allow \emph{multiple} choices, and thus the sum of percentages is greater than 100.) The performance gap between \DeformSyncNet~and Neural Cages (the second best) is statistically significant as the McNemar's contingency test~\cite{McNemar:1947} between the two empirical (binomial) distributions has a p-value of $5e-4$ for the null hypothesis of that the two distributions have an equal marginal. We also remark that for each of the 10 responses collected for each question, \DeformSyncNet~was the most preferred one among all methods in 59 out of 120 questions, and also it tied up with the other most preferred one(s) in 17 questions.

\subsection{Shape Structure Discovery}
\label{sec:structure_discovery}

\begin{figure}[h!]
\centering
\includegraphics[width=0.8\columnwidth]{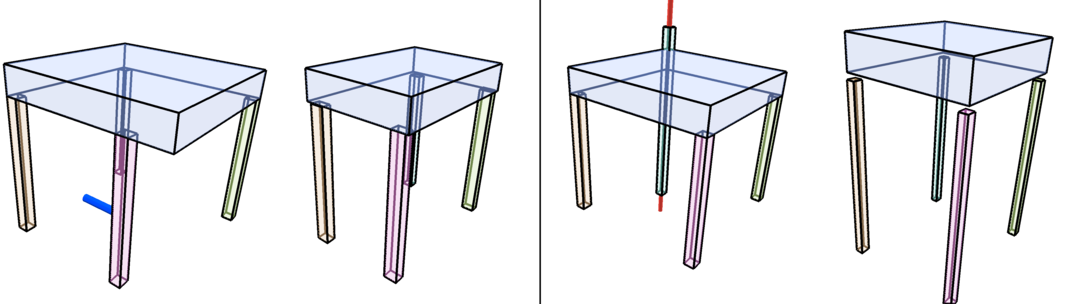}
\vspace{-\baselineskip}
\caption{Examples of perturbed (left) and projected (right) synthetic tables.}
\label{fig:syntable}
\vspace{-0.5\baselineskip}
\end{figure}

We examine the capability of our method of discovering shape structure, such as symmetry and part connectivity. To simplify the test, we employ a procedural model of tables from Tian~\etal~\cite{Tian:2019}, which has a rectangular top and four legs at the corner. Once we train our networks with $5000$ random samples of tables, in test time, we randomly perturb the default shape in $2000$ times and project them back to the learned shape space as described in Section~\ref{sec:qualitative_evaluations}. Figure~\ref{fig:syntable} shows examples of the perturbation and projections. When measuring the difference of symmetric scales, x, y, and z scales of legs, and the gap between the top and legs, the average ratios compared to the scale of the default shape along each axis are $9.60\text{e-}3$, $6.94\text{e-}3$, $9.60\text{e-}3$, and $3.17\text{e-}2$, respectively.

\subsection{Effect of Projection in Training}
\label{sec:effect_of_projection}

\begin{figure}[t!]
\centering
\setlength{\tabcolsep}{0em}
\newcolumntype{Y}{>{\centering\arraybackslash}X}
{\footnotesize
\begin{tabularx}{\columnwidth}{Y|Y|YY|YY}
  \multirow{2}{*}{Source} &
  \multirow{2}{*}{Target} &
  \multicolumn{4}{c}{Deformed} \\
  \cline{3-6}
  & &
  w/o Proj. &
  w/ Proj. &
  w/o Proj. &
  w/ Proj. \\
  \midrule
  \multicolumn{6}{c}{\includegraphics[width=\columnwidth]{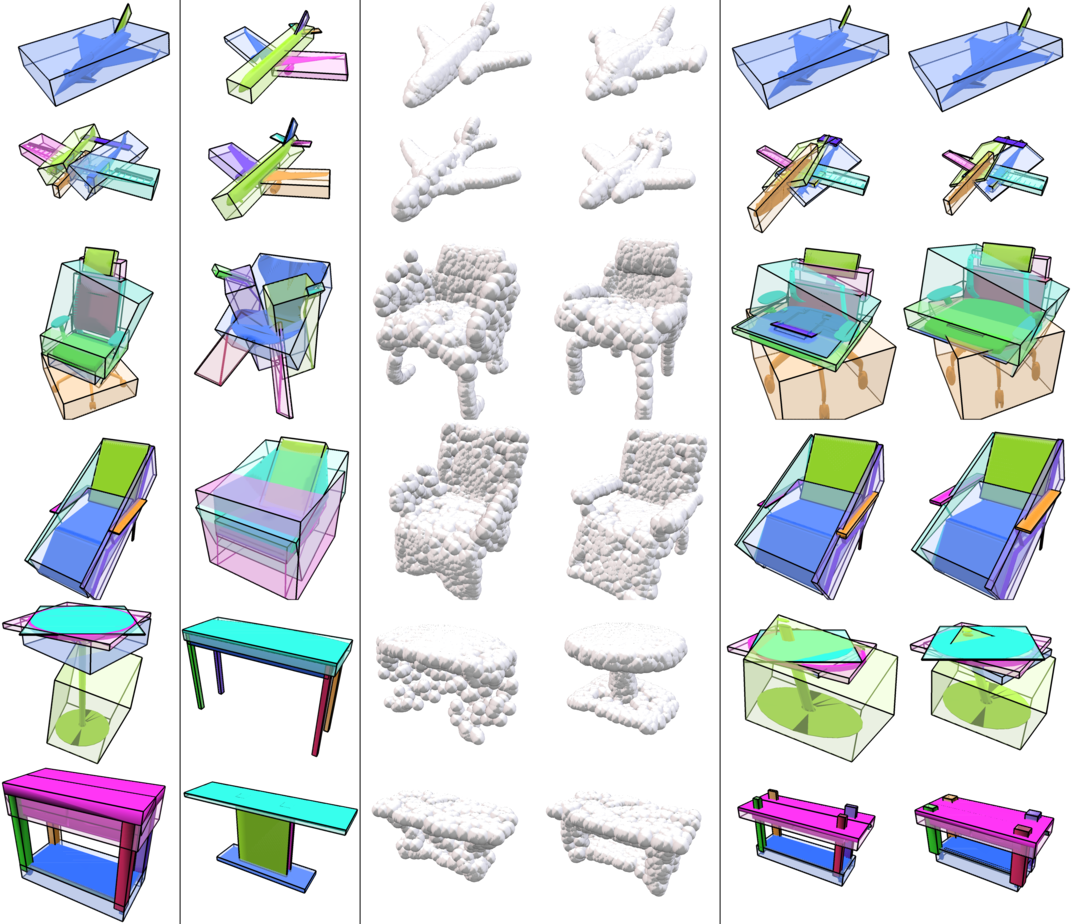}} \\
\end{tabularx}
}
\vspace{-0.5\baselineskip}
\caption{Qualitative analysis of the effect of deformation projection to the deformation handle space during the network training. The third and fourth columns demonstrate that the training with the projection gives more plausible results in deformation. However, if the results are projected in test time, the difference becomes negligible.}
\label{fig:deform_and_transfer_proj}
\end{figure}

When the deformation handles are given for the shapes in the training dataset, we can try to project the deformed source shape to the given deformation handle space during the network training, as explained in Section~\ref{sec:projection}.
\rev{As well as the sparsity loss regularizing the dictionaries (in Equation~\ref{eq:l21_sparsity}), we can also project the deformed shape to the given space (Equation~\ref{eq:deformation_proj}) and measure the fitting loss with it (changing Equation~\ref{eq:fitting_loss} to $\text{Ch} ( d_{\text{proj}}(x \rightarrow y), y )$ ).}
Interestingly, a qualitative analysis demonstrates that the projection during the training guides the network to learn more plausible variations close to the input shape. Figure~\ref{fig:deform_and_transfer_proj} shows some results when training our network with and without the projection during the training using the ShapeNet dataset and part bounding boxes in Section~\ref{sec:qualitative_evaluations}. In test time, the network model trained with the projection provides more reasonable variations of the source shapes (third and fourth columns), while \revfinal{this visual improvement does not make a significant change in the fitting distance to the target} (Table~\ref{tbl:deform_and_transfer_proj}). If we perform the same projection in \emph{test} time (fifth and sixth columns) as well, the difference between two cases, training with and without the projection, becomes negligible despite the discrepancy in the results before the project.

\input{tables/deform_and_transfer_proj.tex}


\subsection{Extension to Non-Linear Deformations}
\label{sec:nonlinear_deformation}
While our framework is designed to decode the latent space to a \emph{linear subspace} of point cloud offsets, we also demonstrate that our framework can be extended to decode each axis of the latent space to a \emph{non-linear} trajectory --- but without guaranteeing the affine properties. The matrix multiplication in Equation~\ref{eq:deformation} can be rewritten as the following \emph{per-point} function:
\begin{align}
    d_i(x \rightarrow y) \coloneqq  \sum_j \{ \mathcal{F}_{ij}(x) \left( \mathcal{E}_j(y) - \mathcal{E}_j(x) \right) \} + x_i,
    \label{eq:deformation_matrix_multiplication}
\end{align}
where $\mathcal{F}_{ij}(x) \in \mathbb{R}^3$ is the $j$-th offset at the $i$-th point, $\mathcal{E}_j(x) \in \mathbb{R}$ is $j$-th element of $\mathcal{E}(x)$, and $x_i$ is the position of $i$-th point of $x$. We  generalize this formulation by redefining $\mathcal{F}_{ij}(x)$ as a \emph{function} describing an \emph{arc-length} trajectory with the parameter $\left( \mathcal{E}_j(y) - \mathcal{E}_j(x) \right)$:
\begin{align}
    d_i(x \rightarrow y) \coloneqq  \sum_j \mathcal{F}_{ij}\left(x, \mathcal{E}_j(y) - \mathcal{E}_j(x) \right) + x_i.
    \label{eq:deformation_nonuniform}
\end{align}

For example, a \emph{uniform circular} trajectory with a parameter $t$ is formulated as follows:
\begin{align}
    \mathcal{F}_{ij}(x, t) \coloneqq \left( \exp \left( [t \mathcal{R}_{ij}(x)]_{\times} \right) - I_{3 \times 3} \right) \left( x_i - \mathcal{C}_{ij}(x) \right),
    \label{eq:deformation_circular}
\end{align}
where $\mathcal{R}_{ij}(x) \in \mathbb{R}^3$ is the rotation vector describing the axis and angle, and $\mathcal{C}_{ij}(x) \in \mathbb{R}^3$ is the rotation center; $[\cdot]_{\times}$ is the cross product matrix of the input vector. This formulation means that each element in the latent vector $\left( \mathcal{E}_j(y) - \mathcal{E}_j(x) \right)$ shared across all the points indicates a scale of the rotation angle or just a rotation angle if $\mathcal{R}_{ij}(x)$ is normalized to a unit vector.

\input{tables/circular_motions.tex}

We test this extension using the Shape2Motion dataset~\cite{Shape2Motion} where the 3D models are annotated with movable parts and their motion parameters, e.g., parameters of rotation and/or translation. We picked three models in different categories, carton, eyeglasses, and swing, which include rotations and generated $1k$ shape variations ($128$ out of them is the test set) by uniformly sampling rotation angles.
Table~\ref{tbl:shape2motion} shows the comparison of fitting errors between the cases of learning linear and circular trajectories (the first two rows). The dimension of the latent space $k$ is set to the degree of freedom of each model. The case learning circular trajectories gives a smaller fitting error. If we set a large number for the dimension of the latent space $k$ (the last two rows), however, the network learning linear trajectories can also provide a very small fitting error by encoding all the rotational motions in the high-dimensional latent space. Note that Neural Cages~\cite{NeuralCages} which also only learns linear deformation offsets fails in this dataset (the fifth row).

\begin{figure}[t!]
\centering
\includegraphics[width=\columnwidth]{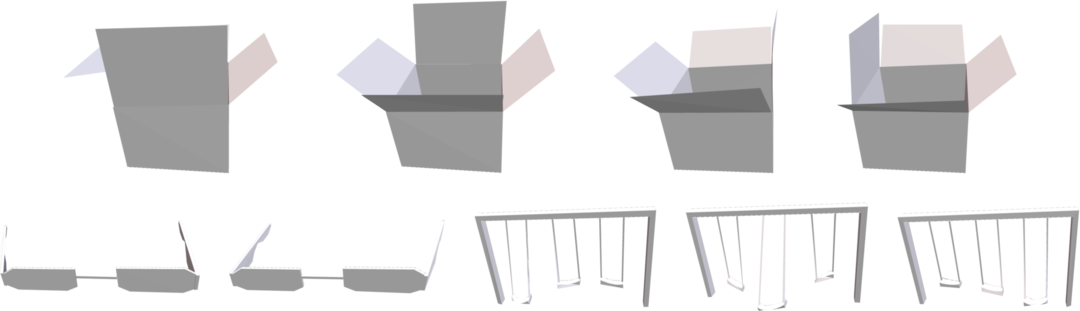}
\vspace{-\baselineskip}
\caption{\rev{The learned dictionaries of \emph{rotational} motions using Shape2Motion dataset~\cite{Shape2Motion}. Each element in the dictionaries describes the motion of each \emph{independent} folding part, such as the flaps of the carton. See supplementary video for animation.}}
\label{fig:shape2motion}
\end{figure}

We also observe that our network can discover independent rotation motions in the learned dictionaries, particularly when the rotation vector $\mathcal{R}_{ij}(x)$ is normalized to a unit vector; this normalization regularizes each rotation element to represent a \emph{same-angle} rotation. Figure~\ref{fig:shape2motion} visualizes the learned rotation dictionaries where each folding part is identified at each element. An animation can also be seen in the supplementary video.

Note that, among the affine properties in Section~\ref{sec:deformation_space}, the transitivity property is not guaranteed when learning the non-linear deformations; i.e., the latent vector may not be uniquely determined given a specific deformation. However, the property can be practically enforced when leveraging regularizations making each element to be unique.

%% file: tables/deform_and_transfer_pc.tex
\begin{table}[t!]
\caption{Comparisons of fitting accuracy of deformation and plausibility of deformation transfer results. We compare our \DeformSyncNet~(\DSN) with Non-rigid ICP (NR-ICP)~\cite{Huang:2017}, 3DN~\cite{3DN}, Cycle Consistency (CC)~\cite{CycleConsistency} and Neural Cages (NC)~\cite{NeuralCages}. Concat$^\ast$ is an ablation study computing the latent deformation vector by concatenating two latent codes of shapes instead of taking the difference. As evaluation metrics for the plausibility and the fitting accuracy of deformation transfer results, we use mean Intersection over Union (mIoU) and Fitting Chamfer Distance (CD) of semantic parts for the former, and Minimum Matching Distance (MMD) and Coverage (Cov) for the latter. MMD and Cov are measured based on Chamfer distance. Bold is the best result, and underscore is the second-best result. See Section~\ref{sec:quantitative_evaluations} for details.}
\vspace{-0.5\baselineskip}
\centering
\newcolumntype{Y}{>{\centering\arraybackslash}X}
{\small
\begin{tabularx}{\columnwidth}{>{\centering}m{1.1cm}|>{\centering}m{0.9cm}|*{5}Y}
\toprule
  \multicolumn{2}{c|}{Category} & Airplane & Car & Chair & Lamp & Table\\
\midrule
  \multirow{5}{*}{\shortstack{mIoU\\(\%)\\$\uparrow$ is better}}
  & NR-ICP & 66.7 & \underline{61.2} & 77.5 & \underline{65.5} & 66.0 \\ 
  & 3DN & 58.4 & 48.3 & 58.1 & 45.9 & 46.6 \\ 
  & CC & \underline{67.8} & 61.1 & 77.6 & \textbf{65.6} & 65.4 \\ 
  & NC & 67.5 & \underline{61.2} & \textbf{78.2} & 64.7 & \textbf{66.9} \\ 
  & \DSN & \textbf{68.0} & \textbf{61.8} & \underline{77.7} & 64.8 & \underline{66.2} \\ 
  & Concat$^\ast$ & 58.6 & 52.9 & 59.1 & 51.7 & 46.6 \\ 
\midrule
  \multirow{5}{*}{\shortstack{Fitting\\CD\\($\times 10^{-3}$)\\$\downarrow$ is better}}
  & NR-ICP & 8.99 & 8.35 & 26.86 & 61.58 & 44.51 \\ 
  & 3DN & 2.54 & 5.04 & 6.32 & 14.15 & 8.51 \\ 
  & CC & 3.26 & \underline{4.18} & 9.81 & 30.65 & 14.61 \\ 
  & NC & 6.73 & 7.49 & 18.82 & 41.40 & 25.80 \\ 
  & \DSN & \underline{1.95} & 4.21 & \underline{5.90} & \underline{13.28} & \underline{8.05} \\ 
  & Concat$^\ast$ & \textbf{1.65} & \textbf{3.98} & \textbf{4.91} & \textbf{9.87} & \textbf{6.48} \\ 
\midrule
  \multirow{5}{*}{\shortstack{MMD-CD\\($\times 10^{-3}$)\\$\downarrow$ is better}}
  & 3DN & 13.31 & 8.15 & 56.18 & 116.95 & 107.14 \\ 
  & CC & 7.44 & 6.45 & 26.83 & 56.89 & 53.39 \\ 
  & NC & \textbf{6.10} & \textbf{4.96} & \underline{21.93} & \underline{52.07} & \underline{33.16} \\ 
  & \DSN & \underline{6.40} & \underline{5.54} & \textbf{21.43} & \textbf{39.36} & \textbf{27.67} \\ 
  & Concat$^\ast$ & 17.85 & 8.84 & 52.45 & 150.40 & 129.40 \\ 
\midrule
  \multirow{5}{*}{\shortstack{Cov-CD\\(\%)\\$\uparrow$ is better}}
  & 3DN & 16.0 & 6.9 & 6.6 & 8.1 & 10.0 \\ 
  & CC & 30.3 & 14.1 & 30.9 & 27.6 & 25.7 \\ 
  & NC & \underline{30.6} & \textbf{37.5} & \underline{39.5} & \underline{32.3} & \underline{35.9} \\ 
  & \DSN & \textbf{31.0} & \underline{33.1} & \textbf{41.3} & \textbf{34.1} & \textbf{38.9} \\ 
  & Concat$^\ast$ & 3.1 & 2.4 & 3.1 & 3.0 & 2.6 \\ 
\bottomrule
\end{tabularx}
}
\label{tbl:deform_and_transfer_pc}
\end{table}

%% file: tables/parallelogram.tex
\begin{table}[t!]
\caption{Quantitative results of parallelogram consistency test. We measure Chamfer distance between two deformation results (going to the same destination but with different pathways): $z \oplus \protect\overrightarrow{x y}$ and $y \oplus \protect\overrightarrow{x z}$. Our \DeformSyncNet~(\DSN) gives the minimal difference between the two results compared with the other methods.}
\vspace{-0.5\baselineskip}
\centering
\newcolumntype{Y}{>{\centering\arraybackslash}X}
{\small
\begin{tabularx}{\columnwidth}{>{\centering}m{1.1cm}|>{\centering}m{0.9cm}|*{5}Y}
\toprule
  \multicolumn{2}{c|}{Category} & Airplane & Car & Chair & Lamp & Table\\
\midrule
  \multirow{5}{*}{\shortstack{CD\\($\times 10^{-3}$)\\$\downarrow$ is better}}
  & 3DN & 13.30 & 7.73 & 77.58 & 194.44 & 133.77 \\ 
  & CC & \underline{6.98} & \underline{5.54} & \underline{26.23} & \underline{46.20} & \underline{57.77} \\ 
  & NC & 8.56 & 7.21 & 31.48 & 138.73 & 89.58 \\ 
  & \DSN & \textbf{2.28} & \textbf{4.28} & \textbf{8.07} & \textbf{22.69} & \textbf{14.33} \\ 
  & Concat$^\ast$ & 16.70 & 7.98 & 69.58 & 103.80 & 122.50 \\ 
\bottomrule
\end{tabularx}
}
\label{tbl:parallelogram}
\end{table}

%% file: tables/deform_and_transfer_proj.tex
\begin{table}[t!]
\caption{Quantitative result of training \DeformSyncNet~with and without the projection to the deformation handle space.
\revfinal{The projection during the training does not make notable changes in the quantities of all the deformation transfer evaluation metrics.}
See Section~\ref{sec:qualitative_evaluations} for the details of the evaluation metrics.}
\vspace{-0.5\baselineskip}
\centering
\newcolumntype{Y}{>{\centering\arraybackslash}X}
{\small
\begin{tabularx}{\columnwidth}{>{\centering}m{0.9cm}|>{\centering}m{1.1cm}|*{5}Y}
\toprule
  \multicolumn{2}{c|}{Category} & Airplane & Car & Chair & Sofa & Table\\
\midrule
  \multirow{2}{*}{\footnotesize{\shortstack{Fitting CD\\($\times 10^{-3}$)$\downarrow$}}}
  & w/o Proj. & 3.53 & \textbf{2.67} & 6.82 & 4.75 & 7.91 \\ 
  & w/ Proj. & \textbf{3.27} & 2.82 & \textbf{5.93} & \textbf{3.76} & \textbf{7.55} \\ 
\midrule
  \multirow{2}{*}{\footnotesize{\shortstack{MMD-CD\\($\times 10^{-3}$)$\downarrow$}}}
  & w/o Proj. & \textbf{1.12} & \textbf{1.97} & \textbf{7.04} & 3.30 & \textbf{6.43} \\ 
  & w/ Proj. & 1.13 & 2.11 & 7.25 & \textbf{3.17} & 6.50 \\ 
\midrule
  \multirow{2}{*}{\footnotesize{\shortstack{Cov-CD\\(\%)$\uparrow$}}}
  & w/o Proj. & \textbf{40.8} & 36.5 & \textbf{39.1} & 38.8 & 40.3 \\ 
  & w/ Proj. & 39.1 & \textbf{37.5} & 38.1 & \textbf{41.2} & \textbf{40.9} \\ 
\bottomrule
\end{tabularx}
}
\label{tbl:deform_and_transfer_proj}
\end{table}

%% file: tables/circular_motions.tex
\begin{table}[t!]
\caption{\rev{Results of learning rotational motions in Shape2Motion dataset~\cite{Shape2Motion} and fitting error comparison. DSN (L) and DSN (C) indicate \DeformSyncNet~learning linear and circular trajectories, respectively. The case learning circular trajectories performs better when the dimension of the latent space $k$ is set to the number of rotating parts (DoF, the first two rows). The case learning linear trajectories also performs well if the dimension of latent space becomes high, e.g., $k=64$ (the last two rows). Neural Cages (NC) fails to learn the rotational motions.}}
\vspace{-0.5\baselineskip}
\centering
\newcolumntype{Y}{>{\centering\arraybackslash}X}
{\small
\begin{tabularx}{\columnwidth}{c|c|*{3}Y}
\toprule
  \multicolumn{2}{c|}{Model} & Carton & Eyeglasses & Swing\\
\midrule
  \multirow{7}{*}{\shortstack{CD\\($\times 10^{-3}$)\\$\downarrow$ is better}}
  & \DSN~(L,$k=$DoF)& 0.14 & 0.33 & 0.41 \\
  & \DSN~(C,$k=$DoF)& \textbf{0.12} & \textbf{0.20} & \textbf{0.32} \\
\cline{2-5}
  & 3DN             & 0.04 & 0.08 & \textbf{0.06}\\
  & CC              & 0.09 & 0.26 & 0.78\\
  & NC              & 1.94 & 0.71 & 7.04\\
  & \DSN~(L,$k=64$) & \textbf{0.03} & \textbf{0.03} & 0.09\\
  & \DSN~(C,$k=64$) & \textbf{0.03} & 0.04 & 0.08 \\
\bottomrule
\end{tabularx}
}
\label{tbl:shape2motion}
\end{table}

%% file: sections/conclusion.tex

\section{Conclusion and Future Work}
\label{sec:conclusion}
We have proposed \DeformSyncNet, a neural-network-based framework learning a synchronized linear deformation space for each shape. The synchronization is achieved without supervised correspondences but by connecting each shape-specific deformation space with an idealized canonical latent space, where all possible deformations are encoded. From this latent space, an encoded deformation is realized directly on each shape through per-point offsets via a shape-specific action decoding. As applications, our framework demonstrates (i)~deformation projection, snapping an edited shape by the user to plausible shape space, and (ii)~deformation transfer, adopting the modification performed on one shape to other shapes in the same category.

Our framework has several limitations.
\rev{While we leverage the deformation handle information during the network training via projection, it is only exploited in the loss function but not fed as input to the network.}
Since most of the deformation handles are associated with not the whole but a part of the shape, part-level features related to them can provide additional information for the deformation space of the shape. Also, we take point clouds as input and use Chamfer distance as the only supervision. A more advanced backbone architecture and regularization losses for handling meshes can help learn more plausible deformations~\cite{Gao:2018,3DN}.
\rev{We also introduced the extension of our framework to non-linear deformation in Section~\ref{sec:nonlinear_deformation}, but enforcing the affine properties remains to be explored.}

Furthermore, the variations of \emph{articulated} shapes may include \emph{hierarchical} structure, such that the variation of a smaller part is factorized from the variation of a larger part. Such a structure might be better understood by finding elements in the deformation dictionary not in parallel, but sequentially. Finally, analogies, i.e., a  transfer of the difference between two shapes to a third, can be extended to cross-domain cases, such as 3D shapes from/to images~\cite{Mo:2019:StructureNet} or natural language~\cite{ShapeGlot:2019}.